\documentclass[reprint,aps]{revtex4-1}
\usepackage{amsfonts}
\usepackage{amssymb}
\usepackage{amsmath}
\usepackage{bbm}
\usepackage{bm}
\usepackage{relsize}
\usepackage{graphicx}

\DeclareMathAlphabet\mathbfcal{OMS}{cmsy}{b}{n}

\newcommand{\be}{\begin{equation}}
\newcommand{\ee}{\end{equation}}
\newcommand{\bea}{\begin{eqnarray}}
\newcommand{\eea}{\end{eqnarray}}
\newcommand{\Eq}[1]{Eq.\,(\ref{#1})}
\newcommand{\Eqs}[2]{Eqs.\,(\ref{#1}) and (\ref{#2})}
\newcommand{\Eqsss}[3]{Eqs.\,(\ref{#1}), (\ref{#2}), and (\ref{#3})}

\newcommand{\Fig}[1]{Fig.\,\ref{#1}}
\newcommand{\Figs}[2]{Figs.\,\ref{#1} and \ref{#2}}
\newcommand{\Sec}[1]{Sec.\,\ref{#1}}
\newcommand{\Secs}[2]{Secs.\,\ref{#1} and \ref{#2}}
\newcommand{\App}[1]{Appendix\,\ref{#1}}

\newcommand\wI{\hat{\mathbb I}}
\newcommand\wG{\hat{\mathbb G}}
\newcommand\wM{\hat{\mathbb M}}
\newcommand\wP{\hat{\mathbb P}}

\newcommand\wD{\hat{\mathbb D}}

\newcommand\wF{\vec{\mathbb F}}

\newcommand\one{\hat{\mathbf{1}}}

\newcommand{\lam}{\lambda}
\newcommand{\eps}{\varepsilon}
\newcommand{\heps}{\hat{{\pmb{\varepsilon}}}}
\newcommand{\hmu}{\hat{{\pmb{\mu}}}}
\renewcommand{\r}{\textbf{r}}
\newcommand{\Y}{\textbf{Y}}
\newcommand{\E}{\textbf{E}}

\newcommand{\D}{\textbf{D}}
\renewcommand{\H}{\textbf{H}}
\newcommand{\B}{\textbf{B}}

\newcommand{\Ec}{\mathbfcal{F}}
\newcommand{\cE}{\mathcal{E}}
\newcommand{\cK}{\mathcal{K}}
\newcommand{\cM}{\mathcal{N}}
\newcommand{\cW}{\mathcal{W}}
\newcommand{\cA}{\mathcal{A}}

\newcommand{\cH}{\mathcal{H}}
\newcommand{\cG}{\mathcal{G}}
\newcommand{\cR}{\mathcal{R}}

\newcommand{\hR}{\hat{\mathcal{R}}}
\newcommand{\Pc}{\hat{\mathcal{P}}}
\newcommand{\Dc}{\hat{\mathcal{D}}}
\newcommand{\cV}{{V}}

\newcommand{\hnabla}{\hat{\nabla}}

\newcommand{\GF}{\hat{\mathcal{G}}}

\newcommand{\hG}{\hat{G}}
\newcommand{\hL}{\hat{L}}
\newcommand{\hM}{\hat{\mathcal{M}}}
\newcommand{\hO}{\hat{O}}
\newcommand{\hQ}{\hat{Q}}

\newcommand{\tV}{\widetilde{V}}
\newcommand{\hcL}{\hat{\cal L}}
\newcommand{\bpsi}{{\pmb{\Psi}}}

\begin{document}

\title{Full electromagnetic Green's dyadic of spherically symmetric open optical systems and elimination of static modes from the resonant-state expansion}

\author{E. A. Muljarov}
\affiliation{%
School of Physics and Astronomy, Cardiff University, Cardiff CF24 3AA, United Kingdom}
\date{\today}

\begin{abstract}

 A general analytic form of the full $6\times6$ dyadic Green's function of a spherically symmetric open optical system is presented, with an explicit solution provided for a homogeneous sphere in vacuum. Different spectral representations of the Green's function are derived using the Mittag-Leffler theorem, and their convergence to the exact solution is analyzed, allowing us to select optimal representations. Based on them, more efficient versions of the resonant-state expansion (RSE) are formulated, with a particular focus on the static mode contribution, including versions of the RSE with a complete elimination of static modes. These general versions of the RSE, applicable to non-spherical optical systems, are verified and illustrated on exactly solvable examples of a dielectric sphere in vacuum with perturbations of its size and refractive index, demonstrating the same level of  convergence to the exact solution for both transverse electric and transverse magnetic polarizations.

\end{abstract}

\maketitle
\section{Introduction}
\label{Sec:Intro}

The electromagnetic dyadic Green's function (GF), introduced by Schwinger more than 70 years ago, is a tensor determining the electric and magnetic fields generated by a point-like source, such as a dipole, an oscillating charge, or a current.
The GF contains a complete information about the physical system and provides access to any observable, such as electromagnetic near and far field distributions~\cite{Tai71,Chew95}, total radiation intensity and
Purcell's factor~\cite{DungPRA00,MuljarovPRB16Purcell}, optical scattering matrix and scattering
cross sections~\cite{LobanovPRA18,WeissPRB18}.

In free space, the Green's dyadic has a closed analytic form~\cite{LevineCPAM50}, clearly demonstrating its spatial singularity. This singularity has a fundamental origin related to the vectorial nature of the electromagnetic field and corresponds to the zero-frequency, i.e. static pole of the GF in the complex frequency plane, responsible for the longitudinal components of the fields. In optical systems, this static pole singularity  can be strongly modified by spatial inhomogeneities of the permittivity and permeability, which presents a significant challenge for its correct calculation.
A comprehensive analysis of the dyadic GFs in electromagnetic systems, including their expansion in bounded media in terms of electric and magnetic eigenmodes of optical resonators and waveguides, was presented in~\cite{Tai71}. Taking into account only the physical modes (which are solenoidal in nature), this treatment, however, was lacking completeness necessary for a correct description of the static-pole singularity. Later on, this mistake was fixed~\cite{TaiIEEE73,CollinCJP73} by adding longitudinal modes to the eigenmode expansion of the dyadic GF~\cite{CollinEM86}.
Still, the static pole problem has caused long debates in the literature~\cite{JohnsonRS79,WangIEEE82} and further attempts to express the GF only in terms of the solenoidal fields~\cite{PathakIEEE83}.

A more analytical approach to the dyadic GF of an open system was developed in the spirit of the scattering Mie theory~\cite{MieAP08,Bohren1998}, by using spherical transverse functions ${\bf M}$ and ${\bf N}$, and longitudinal functions ${\bf L}$, originally introduced by Stratton~\cite{Stratton41}. This approach is based on the assumption of  homogeneity of a spherically symmetric system in the radial direction. Therefore, it has become a rather standard way of treating homogeneous systems~\cite{Chew95} which was intensively used e.g. for multilayered spherical systems~\cite{WeiIEEE94,OkhmatovskiIEEE03,KimMOTL07,FallahiIEEE11}.
There was even an attempt to generalize this formalism for radially inhomogeneous systems~\cite{YehPR63}; however, it is not clear what are the practical benefits of the suggested generalization.

In Stratton's theory, the static pole of the GF of a spherically symmetric multilayered system is build up with ${\bf L}$ functions leading to rather simple analytic expressions~\cite{Chew95}. However, there is also a significant implicit contribution to the static pole coming from the transverse functions ${\bf M}$ and ${\bf N}$.
It is not obvious whether or not this approach treats the static pole of the GF correctly, as no reliable checks of the basis completeness have been performed, to the best of our knowledge. In fact, the analytic results available for the far field (including the Mie theory itself) do not contain any contribution of static modes~\cite{Stratton41,Bohren1998}. It is known, however, that static modes do contribute to the near field~\cite{LobanovPRA18} and can influence the response of the system to excitations placed in its vicinity.

Investigating the pole structure of the dyadic GF in the complex frequency plane is equivalent to expanding it into the eigenmodes of the optical system. Until recently, such expansions were available only for bounded media, using e.g. Dirichlet boundary conditions~\cite{Tai71,DanieleIEEE84}. For example, the correct GF of a {\em closed} spherical cavity and its expansion into the eigenmodes, with a proper account of its static pole in terms of the longitudinal modes, was  presented in~\cite{CollinEM86}. At the same time, similar expansions of the dyadic GFs for {\em open} systems
were not available in electrodynamics. In non-relativistic quantum mechanics dealing with scalar GFs, such expansions are known as Mittag-Leffler (ML) representations~\cite{MorePRA71,MorePRA73}. The major obstacle for applying the same principle to electrodynamics was the normalization of the electromagnetic modes of an open system which was not known. As a result, even for a homogeneous dielectric sphere in vacuum, a proper ML representation of the GF is still missing in the literature.

The electromagnetic modes of an open optical system, called resonant states (RSs) are discrete solutions of Maxwell's equations with outgoing boundary conditions. The RS frequencies are generally complex, reflecting the fact that the energy leaks out of the system. In particular, the quality factor of a RS is given by half of the ratio of real to imaginary part of its eigenfrequency. The concept of RSs has recently become a powerful tool widely used in the literature for studying the spectral properties of open optical systems and for describing resonances observed in the optical spectra in a mathematically rigorous
way~\cite{MuljarovEPL10,DoostPRA13,SauvanPRL13,BaiOE13,DoostPRA14,MuljarovPRB16Purcell,
PerrinOE16,MuljarovPRB16,WeissPRL16,WeissPRB17,LobanovPRA17,LobanovPRA18,WeissPRB18,
MuljarovOL18,BurgerPRA18,YanPRB18,Lalanne18,LalanneJOSAA19,SehmiPRB20,Sam19}.

Finite quality factors of the RSs, while reflecting a leakage of the electromagnetic energy contained within the system to the exterior, also lead to a catastrophic spatial divergence of the RS wave functions. As a result, the standard normalization, given by the volume integral of the square modulus of the wave function, is no longer applicable.
Only recently, the correct normalization of the electromagnetic RSs has been found~\cite{MuljarovEPL10} providing a general analytic expression for an arbitrary dielectric system, which was later on generalized to systems with frequency dispersion~\cite{MuljarovPRB16,MuljarovPRB16Purcell} and arbitrary permeability and chirality~\cite{MuljarovOL18}.

On the other hand, in a purely numerical approach to the RS normalization developed in~\cite{SauvanPRL13}, the exponential growths of the RS fields is damped by introducing so-called perfectly matched layers, artificially absorbing the diverging electromagnetic field and in this way approximating the actual open physical system with an effective closed one. This approach is using a phenomenological expansion of the dyadic GF into a few dominant eigenmodes of the effective closed system. Later on, the method was refined~\cite{YanPRB18} by taking into account in the GF expansion more eigenmodes, including a large number of non-physical states of the absorbing layer, which were required for completeness. Alternative numerical approaches to the normalization and spectral representation of the GF
have been also suggested~\cite{BaiOE13,PerrinOE16,BurgerPRA18}.
In particular, a Riesz-projection method, developed in~\cite{BurgerPRA18} for an efficient treatment of optical systems in terms of only a few RSs close to the frequency range of interest, does not require any explicit mode normalization. It introduces a finite closed contour in the complex frequency plane, and numerically evaluates the contour integral, which can be  understood as a modified ML representation for a limited number of RSs. A more detailed literature review of modern theoretical and computational methods based on the use of the RSs can be found e.g. in \cite{Lalanne18,LalanneJOSAA19,SehmiPRB20}.

Following the analytical approach to scalar GFs developed in quantum mechanics~\cite{MorePRA71,MorePRA73} and using some general properties of GFs in one dimension~\cite{Morse53}, a rigorous ML representation of the electromagnetic GF of a homogeneous dielectric sphere in vacuum was presented in~\cite{MuljarovEPL10} for transverse electric polarization, also verifying the general analytic normalization of the RSs introduced in that work.
Strictly speaking, the ML representation defines the RS normalization via the residues at the poles of the GF, which are located in the complex frequency plane exactly at the RS eigen frequencies. This allowed us to work our later on a rigorous proof of the general analytic normalization of the RSs of an arbitrary three-dimensional (3D) open optical system~\cite{DoostPRA14} and to develop its further generalization~\cite{MuljarovPRB16,MuljarovPRB16Purcell,MuljarovOL18} and application to various geometries~\cite{DoostPRA13,WeissPRL16,WeissPRB17,LobanovPRA17,Sam19}. As a result, a ML representation of the dyadic GF of an arbitrary optical system was obtained~\cite{MuljarovOL18}. This form contains a summation over all the RSs of the system, supplemented with a proper set of static modes required for completeness~\cite{LobanovPRA19}.


The benefit of using the ML representation of the GF is not only that it reveals the pole structure of the Green's dyadic. It also provides the fastest calculation of the optical spectra, as it addresses all the driving frequencies simultaneously. In fact, the optical spectra are given in the form of a superposition of complex Lorentzian lines, each line due to an individual RS. Examples available in the literature include but are not limited to the exact theory of the Purcell effect~\cite{MuljarovPRB16Purcell},  scattering cross-section of micro- and nano-particles~\cite{LobanovPRA18}, and scattering matrix of planar optical systems~\cite{WeissPRB18,WeissPRL16,WeissPRB17}.

The ML representation of the GF is also at the heart of the resonant-state expansion (RSE), a novel rigorous method developed in~\cite{MuljarovEPL10} for calculating the RSs of an arbitrary open optical system. The RSE maps the set of Maxwell's equations onto a linear matrix eigenvalue problem, using the RSs of another system as a basis for expansion.
The basis system differs from the target system by a perturbation and is usually (but not necessarily~\cite{WeissPRL16,WeissPRB17}) solvable analytically. In three dimensions, a homogeneous sphere in vacuum is obviously the simplest basis system allowing an exact analytic solution. It is important to note that the RSE is not limited to small perturbations but is capable of treating perturbations of arbitrary strength, and can be superior to existing computational methods in electrodynamics, such as finite difference in time domain  and finite element methods, in terms of accuracy and efficiency, as demonstrated in~\cite{DoostPRA14,LobanovPRA17,LobanovPRA19}. Another significant advantage of the RSE compared to other methods is that it calculates an asymptotically complete set of the RSs of the target system within a wide spectral range; no RSs are missing and no spurious solutions are produced. The technical implementation is also very straightforward, as the RSs of the target system are found by just diagonalizing a complex matrix containing the matrix elements of the perturbation. Finally, the RSE is a numerically exact method: The only parameter of the RSE is the size of the truncated basis which can be made arbitrarily large.

While applying the RSE to 3D open optical systems, it turned out that in addition to the RSs of the basis system, one has to include in the basis for completeness also an additional sets of static modes, in this way representing the static pole of the GF discussed above. In spite of the fact that the problem of static modes in the RSE has been addressed in~\cite{DoostPRA14,LobanovPRA19}, the RSE method still has an unsolved fundamental problem of correct and efficient inclusion of static modes, or even their partial or complete elimination. Indeed, there is presently available either a quick but incomplete static mode inclusion~\cite{DoostPRA14}, or a complete inclusion of static modes which however suffers from a too slow convergence to the exact solution~\cite{LobanovPRA19}. 

The purpose of the present paper is two-fold: (i) to derive explicit analytic expressions for the dyadic GF of a spherically symmetric system and to find its ML representations properly describing the static pole of the GF, and (ii) to address the static-mode challenge of the RSE, by developing new exact and quickly convergent versions of the method.

As for the {\it first aim}, the general analytic form of the full dyadic GF of an arbitrary spherically symmetric system,
which we derive in this paper, has not been presented in the literature, to the best of our knowledge.
In particular, the provided solution has a number of important features which have not been addressed.

First of all, instead of using the widely applied Stratton's functions ${\bf M}$, ${\bf N}$, and ${\bf L}$, having a specific radial dependence, we implement the formalism of vector spherical harmonics (VSHs)~\cite{BarreraEJP85}. These do not depend on the radial coordinate and are thus suited for treating any radial inhomogeneity. The basis of VSHs provides an elegant mapping of Maxwell's equations onto a 1st-order matrix differential equation describing the radial dependence of the fields. This formalism is useful also for non-spherical systems, as in the far field any solution naturally splits into spherical waves described by the VSHs. Importantly, the latter present a useful basis for calculating the light scattering~\cite{LobanovPRA18,WeissPRB18}.

Secondly, the electromagnetic Green's dyadic is defined in the literature as either electric or, rarely, magnetic Green's tensor of Maxwell's wave equation for, respectively, the electric or magnetic field. Only recently, the full electromagnetic dyadic GF for the set of Maxwell's equations was introduced in \cite{MuljarovOL18}, with both electric and magnetic components contributing on equal footing. Following~\cite{MuljarovOL18}, we treat here the full $6\times6$ Green's tensor satisfying the first-order Maxwell equations with point-like source terms.

Thirdly, for spherically symmetric systems, we obtain a general analytic form of the full dyadic GF, after splitting it into two orthogonal polarizations, transverse electric (TE) and transverse magnetic (TM). Furthermore, we analyze the pole structure of the dyadic GF in the complex frequency plane and derive ML representations properly treating the static pole. Finally, we derive explicit analytic expressions for the GF of a homogeneous sphere in vacuum, even though this solution is available in the literature in some form~\cite{Chew95,KaliteevskiPRB01,ChoPRB02,GlazovPSS11}. We emphasize, however, that a valid ML representation of the GF of a sphere and in particular a correct treatment of its static pole is still missing. Since the RSE is normally using a homogeneous sphere as a basis system, it is very important to know the correct analytic form and a proper ML representation of its dyadic GF.

The correct treatment of the static pole of the dyadic GF is one of the main achievements of the present work. Based on this knowledge, the full ML expansion of the GF is presented in several different ways.
Different ML representations can also lead to different versions of the RSE.
This flexibility is due to the energetic degeneracy of static modes, so that one can use any suited basis in order to represent the static pole of the GF, including a basis build up from the RSs themselves. In the latter case, static modes are effectively eliminated from the basis. Such an elimination of static modes from the RSE basis and a linked to it task  of RSE optimization are the {\it second aim} and the main focus of the present work.

The paper solves this optimization problem by considering four different ML representations of the GF of the basis system and following from them four different versions of the RSE. The presented theory is general and suited for arbitrary 3D open optical systems treated by the RSE. The limitation to spherically symmetric systems is related to the properties of the basis system only. However, as required for verification, illustrations of the new versions of the RSE are provided for the exactly solvable case of an ideal sphere in vacuum. Comparisons with available commercial solvers treating non-spherical cases numerically, similar to those provided in~\cite{DoostPRA14,LobanovPRA19}, will be published elsewhere. Studying the convergence of the RSE towards the available exact solutions, we find the optimal versions of the method which can further be tested and used for non-spherical perturbations.

The paper is organized as follows.  In~\Sec{Sec:RSs} we first briefly summarize the existing theory of the RSs, providing known results for their normalization, orthogonality, and completeness, and based on these properties, a ML representation of the dyadic GF of an arbitrary finite optical system, including the contribution of static modes. The standard version of the RSE available in the literature is then presented in~\Sec{Sec:RSE}, with a numerical optimization of the static-mode contribution. We then introduce in~\Sec{Sec:RSE-elim} a new version of the RSE with static modes entirely eliminated from the basis and provide its illustration for a size perturbation of a dielectric sphere in vacuum, demonstrating in particular a slow convergence, very similar to the standard version of the RSE~\cite{LobanovPRA19}.

Section~\ref{Sec:SSS} is devoted to the analytic properties of spherically symmetric systems, described by radially dependent isotropic permittivity and permeability, treated in the basis of VSHs. In~\Sec{Sec:SGF}, the full $6\times 6$ dyadic GF is split into two separate $3\times 3$ blocks, one for TE, the other for TM polarization. Each block is found in terms of scalar solutions of a 2nd-order ordinary differential equation. The static pole of the dyadic GF is studied in~\Sec{Sec:StaticPole} where it is expressed in terms of a scalar GF, and further expanded into a complete set of static modes in~\Sec{Sec:static}. The RSs of a spherically symmetric system are normalized in~\Sec{Sec:Norm}, which is then used to obtain in~\Sec{Sec:ML} three different ML representations of the dyadic GF, including a regularized, quickly convergent version.  This regularized ML representation is then used in~\Sec{Sec:SRSE} for developing a new efficient version of the RSE.

Sections~\ref{Sec:GFsphere} and \ref{Sec:RSNorm} provide explicit analytic expressions for, respectively, the dyadic GF and normalized RSs of a homogeneous sphere. The analysis of the dyadic GF culminates in~\Sec{Sec:pole} developing a one more ML representation with the static pole expressed in terms of the wave functions of the RSs only. This fourth ML representation provided in the paper is also regular, which results in an efficient variant of the RSE with static modes entirely eliminated from the basis.

The main results of this paper are demonstrated numerically in~\Sec{Sec:Results} using a homogeneous sphere as an exactly solvable system taken for illustration and verification. In particular, convergence of the two developed ML representations, with static mode elimination, towards the analytic solution presented in~\Sec{Sec:GFsphere}, is studied in~\Sec{Sec:NumGF}. The versions of the RSE corresponding to these ML representations are then illustrated in~\Sec{Sec:Shell} on examples of refractive index and size perturbations of the sphere.

Finally, \Sec{Sec:Con} summarizes the main results demonstrated in the paper.
Details of derivations are provided in Appendices~A--C.



\section{Formalism of resonant states in electrodynamics and the resonant-state expansion}
\label{Sec:RSsRSE}

In this section, we first briefly summarize the formalism of the RSs and based on it the RSE for non-dispersive systems, which includes using static modes. We also introduce here a version of the RSE with complete elimination of static modes from the RSE basis.

Let us write, following~\cite{MuljarovOL18}, the set of Maxwell's equations describing electromagnetic waves in a compact symmetric form:
\be
\wM(k,\r) \wF(\r)=0\,,
\label{ME}
\ee
where $k=\omega/c$ is the light wave number,
$$
\wF(\r)= \left(\begin{array}{c}
\E(\r)\\
i\H(\r)\\
\end{array}\right)
$$
is a 6-dimensional vector comprising  the electric field $\E$ and the magnetic field $\H$ on equal footing, and
$$
\wM(k,\r)=k\wP(\r)-\wD(\r)
$$
is a $6\times6$ matrix  Maxwell's operator. The latter consists of a generalized permittivity tensor $\wP(\r)$ and a differential curl operator $\wD(\r)$, which are defined by
\be
\wP(\r)=\left(\begin{array}{cc}
\heps(\r)&0\\
0&\hmu(\r)\\
\end{array}\right)\,,
\ \ \ \
\wD(\r)= \left(\begin{array}{cc}
0&\nabla\times\\
\nabla\times&0\\
\end{array}\right)\,,
\label{perm}
\ee
where $\heps(\r)$ and $\hmu(\r)$ are respectively, the standard $3\times3$ permittivity and permeability tensors which are assumed to be frequency independent.

We next introduce a $6\times6$ generalized dyadic GF $ \wG_k(\r,\r')$ which satisfies
an inhomogeneous equation
\be
\wM(k,\r) \wG_k(\r,\r')=\wI\delta(\r-\r')
\label{GF-equ}
\ee
 and the outgoing boundary conditions for any real $k$ (here $\wI$ is the $6\times6$ identity matrix). The GF satisfies a general reciprocity relation
\be
\wG_k(\r',\r)= \wG_k^{\rm T}(\r,\r')\,,
\label{reciprocity}
\ee
where ${\rm T}$ denotes matrix transposition. This property follows from the reciprocity relations for the generalized permittivity, since  $\heps^{\rm T}=\heps$ and $\hmu^{\rm T}=\hmu$ for any reciprocal medium.

\subsection{Resonant states, static modes, their orthonormality, and Mittag-Leffler series}
\label{Sec:RSs}

The RSs of an optical system are defined as eigen solution of Maxwell's equations,
\be
\wM(k_n,\r) \wF_n(\r)=0\,,
\label{RS-equ}
\ee
satisfying outgoing wave boundary conditions. Here, $k_n$ is the RSs eigen wave number, and index $n$ is used to label the RSs.

Strictly speaking, purely outgoing waves can be observed only for a real $k$, e.g. in the GF. At the same time, the wave functions of the RSs with Re\,$k_n<0$ and small negative imaginary part of $k_n$ are looking like incoming-wave solutions. Nevertheless, they contribute to the GF satisfying the outgoing wave boundary conditions, and therefore are formally classified as eigen solutions with outgoing waves outside the system.

In addition to the RSs, all having non-vanishing complex eigen wave numbers $k_n$, there are also zero frequency ($k=0$), static solutions of Maxwell's equations (\ref{ME}). The latter take the following form in the static limit:
\be
\begin{array}{c}
\nabla \times \E_\lam(\r)=0\,,
\smallskip
\\
\nabla \times \H_\lam(\r)=0\,.
\end{array}
\label{static}
\ee
Here, static modes are labeled with index $\lambda$. Note that both lines in \Eq{static} are independent of each other, so that static electric and static magnetic modes can be considered as two separate groups of modes. Each group is represented by longitudinal  fields,
\be
\begin{array}{lll}
\E_\lam=-\nabla \psi^{\rm LE}_\lam\,, & \H_\lam=0 & ({\rm electric}),
\smallskip
\\
\E_\lam=0\,, & \H_\lam=-\nabla \psi^{\rm LM}_\lam  & ({\rm magnetic}),
\end{array}
\label{LELM}
\ee
expressed in terms of scalar potentials $\psi^{\rm LE}_\lam(\r)$ and $\psi^{\rm LM}_\lam(\r)$ for, respectively, longitudinal electric (LE) and longitudinal magnetic (LM) modes.

For the RSs, as they all have $k_n\neq 0$, the other pair of Maxwell's equations,
\be
\begin{array}{c}
\nabla\cdot \D_n=0\,,
\smallskip
\\
\nabla\cdot \B_n=0\,,
\end{array}
\label{ME2}
\ee
where $\D_n=\heps \E_n$ and $\B_n=\hmu H_n$, is satisfied automatically,  as it follows from \Eq{RS-equ}. For static modes, however, fulfilling \Eq{ME2} is not guaranteed. This determines the nature of static modes, potentially carrying volume and surface charged as it has been discussed in detail in~\cite{LobanovPRA19}. This property of static modes and their degeneracy with respect to the wave number bring in some uncertainty, or rather, a degree of freedom for their inclusion into the ML form of the GF and the RSE. In fact, the full dyadic GF $ \wG_k(\r,\r')$ contains a $k=0$ pole which originates from the longitudinal divergent part of the electromagnetic free-space dyadic GF~\cite{LevineCPAM50}. The pole residue modifies in the presence of inhomogeneities. However, its singular part remains the same.  This pole corresponds to and can be described with static solutions satisfying \Eq{static}. Having an infinite-multiple degeneracy (unlike the poles of the GF due to the RSs which can only have finite degeneracy by symmetry or due to exceptional points~\cite{WiersigPRA11}), this pole presents a significant challenge in applying the ML theorem to the dyadic GF, which is tackled in the present work.

Now, adding to the full set of the RSs of the optical system any complete set of its static modes, we obtain a spectral representation of the dyadic GF
\be
\wG_k(\r,\r')=\sum_\nu \frac{\wF_\nu(\r)\otimes \wF_\nu(\r')}{k-k_\nu}\,,
\label{GF-ML}
\ee
valid at least within a minimal convex volume including the system.
Equation (\ref{GF-ML}) follows from applying the ML theorem to the GF and using its reciprocity~\cite{DoostPRA13,MuljarovOL18}.
Here, index $\nu$ is introduced for convenience to label together all
the RSs and static modes contributing to the ML series~\Eq{GF-ML}.
However, in each group, modes have their own labels: index $n$ is used throughout this paper for RSs only and $\lambda$ for static modes only.  $\otimes$ denotes the dyadic product of vectors.
The ML form \Eq{GF-ML} defines the normalization of electromagnetic modes~\cite{MuljarovEPL10,DoostPRA14,MuljarovPRB16Purcell,MuljarovOL18}, which can be written for the RSs as
\bea
1&=&\int_\cV \left( \E_n\cdot \heps\E_n- \H_n\cdot \hmu \H_n\right) d\r
\label{Norm}\\
&&+\frac{i}{k_n}\oint_{S_\cV}\left[\E_n\times (\r\cdot\nabla)\H_n+\H_n\times (\r\cdot\nabla)\E_n\right]\cdot d{\bf S}\,,
\nonumber
\eea
where $V$ is an arbitrary volume containing all the system inhomogeneities and $S_\cV$ is its boundary. For static modes, the normalization reduces to
$$
1=\int \left( \E_\lam\cdot \heps\E_\lam- \H_\lam\cdot \hmu \H_\lam\right) d\r
$$
with the integral extended to the full space, owing to the square integrable wave functions of the static modes~\cite{DoostPRA14,LobanovPRA19}.

The orthogonality of the RSs in turn follows directly from Maxwell's equations (\ref{RS-equ}) and has a similar form~\cite{MuljarovEPL10,DoostPRA14}
\bea
0&=&(k_\nu-k_{\nu'}) \int_\cV \left( \E_\nu\cdot \heps\E_{\nu'}- \H_\nu\cdot \hmu \H_{\nu'}\right) d\r
\nonumber\\
&&+i\oint_{S_\cV}\left(\E_{\nu}\times \H_{\nu'}+\H_{\nu}\times \E_{\nu'}\right)\cdot d{\bf S}\,,
\nonumber
\eea
valid for $k_\nu\neq k_{\nu'}$. For degenerate modes, the orthogonality is guaranteed by vanishing of the corresponding volume and surface integrals. These integrals vanish by symmetry for degenerate RSs and by both symmetry and orthogonalization of the full-space volume integrals for static modes, see~\cite{LobanovPRA19}.

Substituting the ML expansion \Eq{GF-ML} back into \Eq{GF-equ}, we obtain, using \Eq{RS-equ}, a closure relation
\be
\wP(\r)\sum_\nu \wF_\nu(\r)\otimes \wF_\nu(\r')=\wI\delta(\r-\r')\,,
\label{closure}
\ee
which confirms in particular that the full set of modes is complete, and that any function $\wF(\r)$ within the system volume can be expanded as
\be
\wF(\r)= \sum_\nu c_\nu \wF_\nu(\r)\,.
\label{F-exp}
\ee
In reality, this set is over-complete, so that some reduced subsets of functions can be instead used for expansion, as can be seen in \Sec{Sec:RSE-elim} below.

\subsection{Resonant-state expansion}
\label{Sec:RSE}

Expansions \Eqs{GF-ML}{F-exp} can be used for finding the RSs of a perturbed system, described
by a modified permittivity tensor $\wP(\r)+\Delta\wP(\r)$, where
\be
\Delta\wP(\r)=\left(\begin{array}{cc}
\Delta\heps(\r)&0\\
0&\Delta\hmu(\r)\\
\end{array}\right),
\label{del-perm}
\ee
is a perturbation. The perturbed RSs satisfy Maxwell's equations
\be
[\wM(k,\r)+k\Delta\wP(\r)] \wF(\r)=0\,,
\label{ME-pert}
\ee
and outgoing boundary conditions. Solving \Eq{ME-pert} with the help of the GF of the unperturbed system $\wG_k(\r,\r')$
yields
\bea
\wF(\r)&=&-k\int\wG_k(\r,\r')\Delta\wP(\r')\wF(\r')d\r'\nonumber\\
&=&-k\sum_\nu \frac{\wF_\nu(\r)}{k-k_\nu}\int\wF_\nu(\r')\cdot\Delta\wP(\r')\wF(\r')d\r'\,,
\label{FFF}
\eea
where we have also used the ML expansion \Eq{GF-ML}. Substituting the expansion \Eq{F-exp} into \Eq{FFF} and equating coefficients at $\wF_\nu(\r)$, we arrive at  the RSE matrix equation~\cite{MuljarovEPL10,MuljarovOL18}:
\be
(k-k_\nu) c_\nu = -k \sum_{\nu'} V_{\nu\nu'} c_{\nu'} \,,
\label{RSE}
\ee
where $k$ is the wave number of a perturbed RS (or a static mode) and $c_\nu$ are the coefficients of the expansion of its wave function $\wF(\r)$ into the unperturbed states $\wF_\nu(\r)$, which is given by \Eq{F-exp}.
The perturbation matrix elements have the following form
\bea
V_{\nu\nu'}&=&\int \wF_\nu(\r)\cdot \Delta\wP(\r)\wF_{\nu'}(\r)d\r\nonumber\\
&=&\int_{\cV_0} \left( \E_\nu\cdot \Delta\heps\E_{\nu'}-\H_\nu\cdot\Delta\hmu \H_{\nu'}\right) d\r\,,
\nonumber
\eea
where $\cV_0$ is the system volume, and the perturbation of the permittivity and/or permeability is assumed to be confined within $\cV_0$. Generalization of this formalism to systems with frequency dispersion is provided in~\cite{MuljarovPRB16} and with bi-anisotropy and chirality in~\cite{MuljarovOL18}.

It is beneficial for numerical efficiency of solving \Eq{RSE} to separate the RS and the static mode contributions, by writing \Eq{F-exp} as
$$
\wF(\r)= \sum_n c_n \wF_n(\r)+ \sum_\lam c_\lam \wF_\lam(\r)\,,
$$
where indices $n$ and $\lambda$ label the RSs and static modes, respectively.
Owing to the degeneracy of static modes, the RSE equation (\ref{RSE}) can be reduced to a linear matrix eigenvalue problem formulated in terms of the basis RSs only~\cite{LobanovPRA19}
\be
(k-k_n) c_n = -k \sum_{n'} \tV_{nn'} c_{n'}\,,
\label{RSE-mod1}
\ee
where
$$
 \tV_{nn'}= V_{nn'}-\sum_{\lam \lam'}V_{n\lam} W_{\lam\lam'} V_{\lam' n'}
$$
and $W_{\lam\lam'}$ is the inverse of matrix $\delta_{\lam\lam'}+V_{\lam\lam'}$.  The static-mode coefficients $c_\lam$ are given by
$$
c_\lam = - \sum_{\lam'} W_{\lam\lam'} \sum_{n} V_{\lam' n} c_{n} \,.
$$
The numerical procedure can be further optimized
by introducing new coefficients~\cite{MuljarovEPL10}
$$
b_n=\sqrt{\frac{k_n}{k}} c_n\,.
$$
Then the RSE equation (\ref{RSE-mod1}) reduces to diagonalization of a complex symmetric matrix:
\be
\sum_{n'} \left(\frac{\delta_{nn'}}{k_n}+\frac{\tV_{nn'}}{\sqrt{k_n}\sqrt{k_{n'}}}\right)b_{n'}=\frac{1}{k} b_{n}\,.
\label{RSE-mod2}
\ee

Equation~(\ref{RSE-mod1}), equivalent to \Eq{RSE-mod2}, was solved in~\cite{LobanovPRA19} for various perturbations of system's size, shape, and permittivity, including those breaking the spherical symmetry of the basis system. In the example of the size perturbation of a dielectric sphere, in which static modes play a crucial role in calculating the TM modes, solving \Eq{RSE-mod1} demonstrated a very slow, $1/N$ convergence to the exact solution, where $N$ is the basis size. It has been shown in~\cite{LobanovPRA19} that this slow convergence is coming from the static mode contribution. At the same time, the RSE for spherically symmetric perturbations of the TE modes of a homogeneous sphere in vacuum converges to the exact solution as $1/N^3$, since it does not require any static modes~\cite{MuljarovEPL10}.

\subsection{Elimination of static modes}
\label{Sec:RSE-elim}

Instead of numerical exclusion of static modes from the matrix diagonalization problem described in \Sec{Sec:RSE}, one can fully eliminate them on a more fundamental level, by separating the static-mode part of the closure relation \Eq{closure} as
$$
\wP(\r)\sum_\lam \wF_\lam(\r)\otimes \wF_\lam(\r')=\wI\delta(\r-\r')-\wP(\r)\sum_n \wF_n(\r)\otimes \wF_n(\r')
$$
and substituting it into the ML expansion \Eq{GF-ML}. Using the fact that all static modes have $k_\lam=0$, we obtain
\be
\wG_k(\r,\r')=\sum_n \frac{k_n \wF_n(\r)\otimes \wF_n(\r')}{k(k-k_n)}+\frac{1}{k}\wP^{-1}(\r)\delta(\r-\r') \,,
\label{GF-ML2}
\ee
where tensor $\wP^{-1}(\r)$ is the inverse of $\wP(\r)$, and index $n$ labels the RSs only.  We have thus removed any explicit contribution of static modes to the dyadic GF, at the cost of emergence of an additional term with a $\delta$ function. With the help of the new ML form \Eq{GF-ML2}, the solution of the perturbed Maxwell's equations (\ref{ME-pert}) takes the following form
\bea
\wF(\r)&=&-k\int\wG_k(\r,\r')\Delta\wP(\r')\wF(\r')d\r'\nonumber\\
&=&-\sum_n \frac{k_n}{k-k_n} \wF_n(\r)\int\wF_n(\r')\cdot\Delta\wP(\r')\wF(\r')d\r'\nonumber\\
&&-\wP^{-1}(\r)\Delta\wP(\r)\wF(\r)\,,\nonumber
\eea
which can also be written as
\be
\wF(\r)=\sum_n d_n[\wP(\r)+\Delta\wP(\r)]^{-1}\wP(\r)\wF_n(\r)\,,
\label{F-exp3}
\ee
where the coefficients $d_n$ are given by
\be
d_n= -\frac{k_n}{k-k_n} \int\wF_n(\r)\cdot\Delta\wP(\r)\wF(\r)d\r\,.
\label{dn}
\ee

We see that \Eq{F-exp3} is an expansion of a perturbed RS wave function using only the RSs of the unperturbed system, i.e. not involving explicitly any static modes. Substituting \Eq{F-exp3} into \Eq{dn}, we obtain a new RSE equation [compare with \Eq{RSE-mod1}]:
\be
(k-k_n) d_n = -k_n \sum_{n'} U_{nn'} d_{n'} \,,
\label{RSE-d}
\ee
where the matrix elements of the perturbation are now given by
\be
U_{nn'}=\int \wF_n(\r)\cdot \Delta\wP(\r)[\wP(\r)+\Delta\wP(\r)]^{-1}\wP(\r) \wF_{n'}(\r)d\r\,.
\label{Unm}
\ee
Finally, introducing new expansion coefficients
$$
a_n=\sqrt{\frac{k}{k_n}} d_n\,,
$$
the perturbed RSs can be found by diagonalizing another complex symmetric matrix:
\be
\sum_{n'} \left(\delta_{nn'}k_n-U_{nn'}\sqrt{k_n}\sqrt{k_{n'}}\right)a_{n'}=k a_{n}\,.
\label{RSE-mod3}
\ee

Note that the above results are quite general and are valid even if $\wP(\r)$ and/or $\Delta\wP(\r)$ include also bi-anisotropy and chirality tensors. Including the dispersion would modify some of the above results to forms similar to those provided in~\cite{MuljarovPRB16,MuljarovOL18}. For $\wP(\r)$ and $\Delta\wP(\r)$ given by \Eqs{perm}{del-perm}, respectively, the matrix elements \Eq{Unm} take the following explicit form
\bea
U_{nn'}&=& \int  \E_n\cdot \Delta\heps[\heps+\Delta\heps]^{-1}\heps\E_{n'} d\r\nonumber\\
&&-\int\H_n\cdot\Delta\hmu[\hmu+\Delta\hmu]^{-1}\hmu\H_{n'} d\r\,,
\label{Unm-expl}
\eea

Let us consider for illustration a dielectric sphere in vacuum perturbed to a sphere of the same permittivity but a smaller size.   To find perturbed RSs of TM polarization via the standard RSE equation (\ref{RSE}) derived in~\cite{MuljarovEPL10}, one needs to include a complete set of static modes, as it has been done in~\cite{LobanovPRA19}. Here, we use the new version of the RSE, \Eq{RSE-mod3}, with a complete elimination of static modes. Details of the calculation of the unperturbed wave numbers $k_n$ and the matrix elements $U_{nn'}$ can be found in \Sec{Sec:Sphere} and \App{App:Sphere} below, as well as in \cite{DoostPRA14}.

Figure~\ref{Fig:slowRSE}(top) shows the exact values of the unperturbed and perturbed RS wave numbers along with those calculated via the RSE for the size perturbation of the sphere, going from radius $R$ to radius $0.7R$, in this way reducing the whole volume of the sphere by $\sim2/3$. For $N=400$ RSs in the basis, the RSE wave numbers are in visual agreement with the exact values, and the relative error is close to or even less than 1\%, see \Fig{Fig:slowRSE}(bottom). Comparing with the error for 10 times smaller and 10 times larger basis sizes, it becomes clear that the relative error is inversely proportional to the basis size $N$.
	\begin{figure}
\includegraphics*[clip,width=0.48\textwidth]{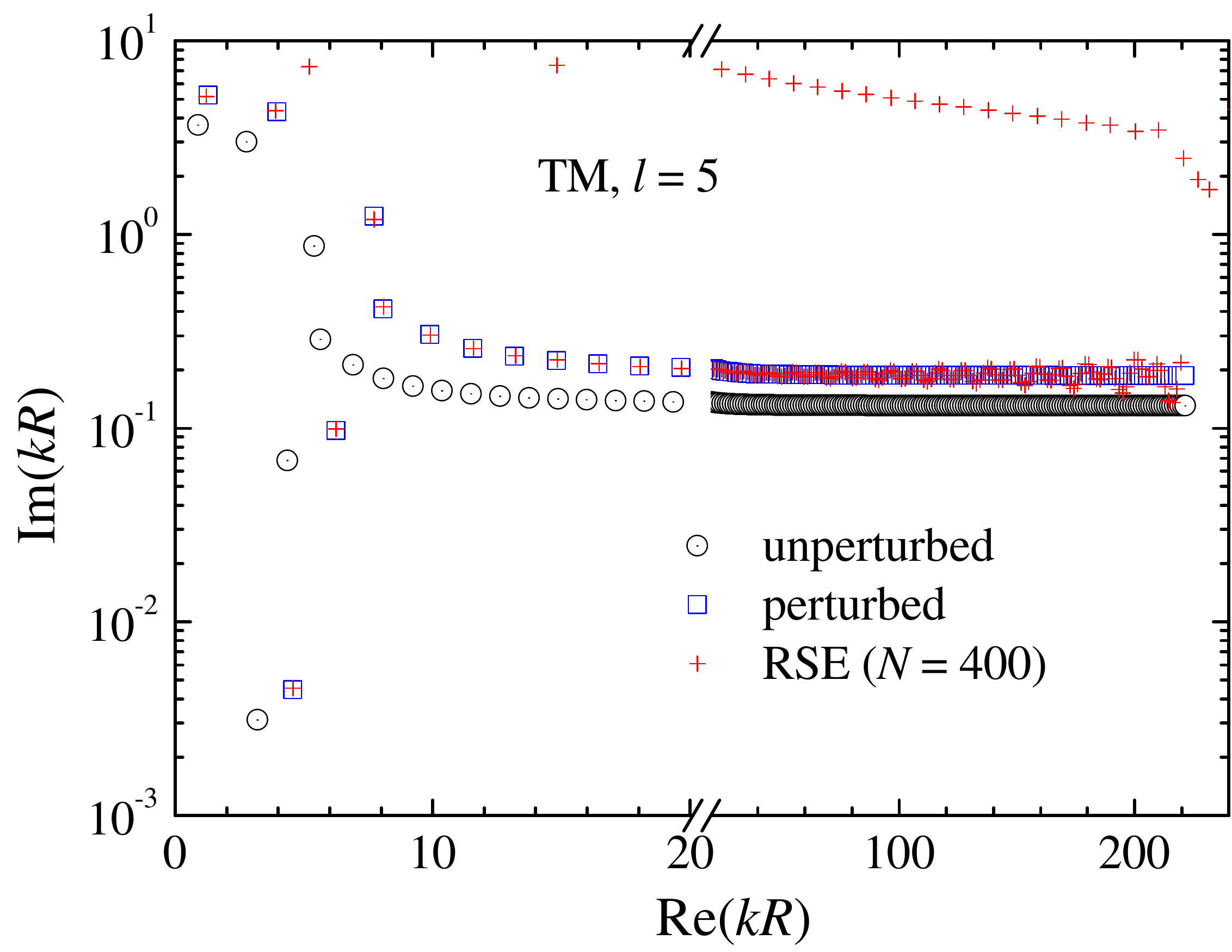}
\includegraphics*[clip,width=0.48\textwidth]{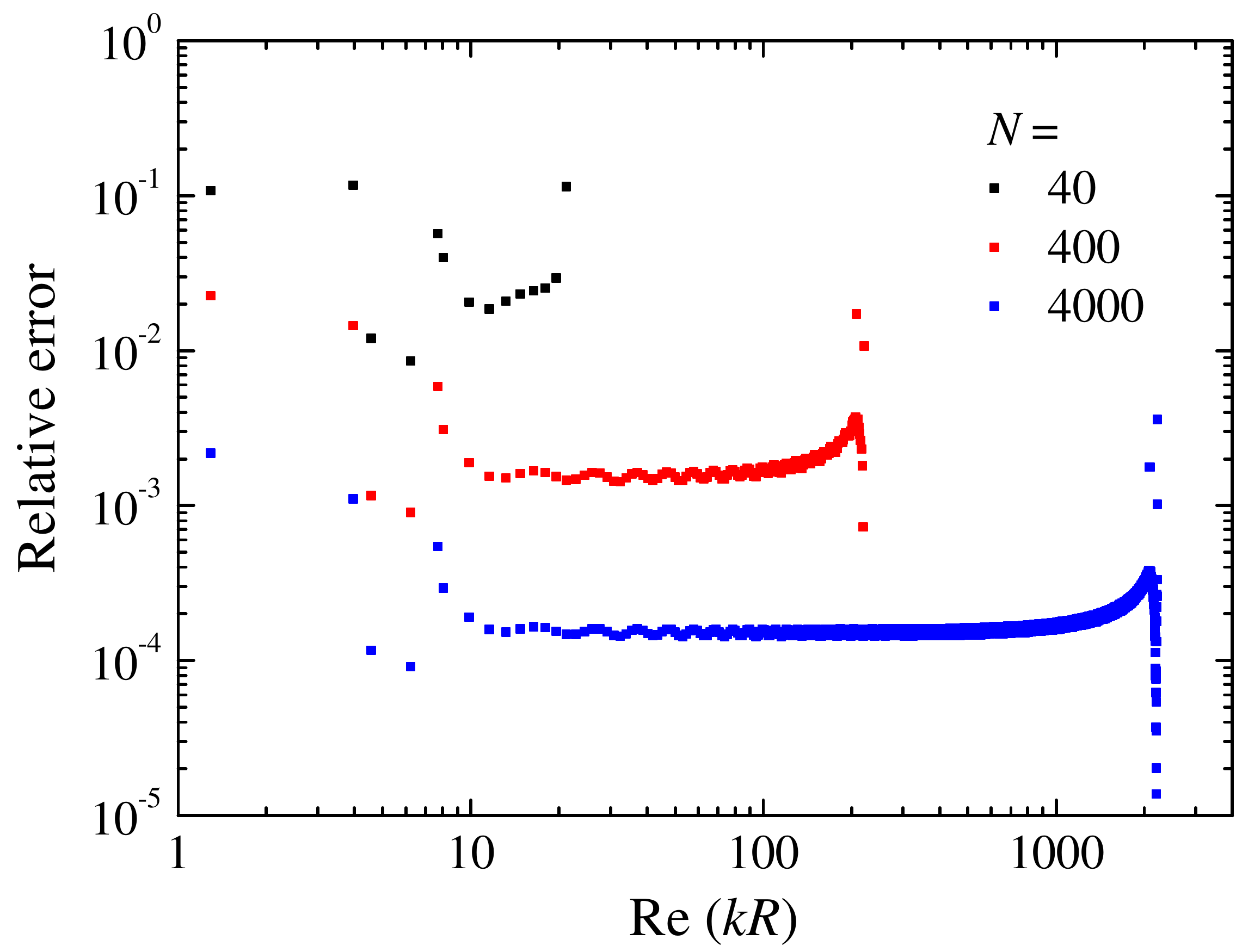}
		\caption{Top: Wave numbers of TM RSs calculated exactly for the unperturbed (black circles with dots) and perturbed system (blue squares), and by solving the RSE equation (\ref{RSE-mod3}) with $N=400$ RSs in the basis (red crosses). The unperturbed (perturbed) system is a homogeneous dielectric sphere in vacuum, with radius $R$ (0.7$R$), permittivity $\eps=8$ and permeability $\mu=1$.  Bottom: Relative error of the RSE calculation of the RS wave numbers for different basis sizes $N$ as given.}
\label{Fig:slowRSE}
	\end{figure}

As we see from this example, the new version of the RSE with complete elimination of static modes, \Eq{RSE-mod3}, has the same slow, $1/N$ convergence to the exact solution as for the standard RSE equation (\ref{RSE-mod1}) [equivalent to (\ref{RSE}) and (\ref{RSE-mod2})] with a full set of static modes included \cite{LobanovPRA19}. The observed poor convergence of these two quite different versions of the RSE, one with and the other without static modes, has provided us with a sufficient motivation for having a closer look at the dyadic GF, focusing in particular on the properties of its $k=0$ pole, and obtaining different representations of the Green's dyadic. This has resulted in developing new and more efficient versions of the RSE having a quicker convergence to the exact solution.

In the following sections we consider rigorously the $k=0$ pole of the dyadic GF of spherically symmetric systems and show that the ML forms \Eqs{GF-ML}{GF-ML2} given above have poor convergence because of the $k=0$ singularity (similar to that in free space~\cite{LevineCPAM50}) represented by a series of smooth functions, which are the wave function of the RSs and/or static modes.
We then work out alternative ML representations of the Green's dyadic and following from them RSE equations which have a much quicker convergence. We also provide in \Sec{Sec:ML}
a rigorous proof of the ML expansions \Eqs{GF-ML}{GF-ML2} for spherically symmetric systems.

\section{Spherically symmetric systems}
\label{Sec:SSS}

We now concentrate on spherically symmetric systems and use the advantage that the full 3D problem for the RSs and the GF in this case can be reduced to effective 1D where many useful properties can be derived analytically. At the same time, we assume in this section an arbitrary radial dependence of the generalized permittivity, thus keeping all the conclusions made in this work as general as possible. We assume that such a spherically symmetric optical system is finite, having radius $R$, and is surrounded by vacuum, although a generalization of the obtained results to arbitrary uniform permittivity of the surrounding medium is straightforward. Application of these results to a homogeneous sphere allowing explicit analytic solutions will be done in the next section.

For a spherically symmetric system, its permittivity and permeability have only radial dependence,
$$
\heps(\r)=\one \eps(r)\,,\ \ \  \hmu(\r)=\one \mu(r)\,,\ \ \
$$
(here, we naturally assume also their isotropy). It is convenient to use vector spherical harmonics (VSHs) $\Y_{jlm}(\Omega)$ ($j=1,2,3$) for solving Maxwell's equations for the RSs and the GF, \Eqs{RS-equ}{GF-equ}, respectively. The VSHs are defined as
\be
\Y_{1lm}=\frac{\r}{\alpha_l} \times \nabla Y_{lm}\,, \ \ \
\Y_{2lm}=\frac{r}{\alpha_l} \nabla Y_{lm}\,, \ \ \
\Y_{3lm}=\frac{\r}{r} Y_{lm}\,,
\label{VSH}
\ee
where
\be
\alpha_l=\sqrt{l(l+1)}\,,
\label{alpha}
\ee
$Y_{lm}(\Omega)$ are scalar spherical harmonics defined in \App{App:VSH}, $l$ and $m$ are the spherical quantum numbers, and $\Omega=(\theta, \varphi)$ is the angular part of the standard spherical coordinates. Using the completeness of the VSHs, we consider an expansion of the electric field into the VSHs:
\be
\E(\r)=\sum_{jlm} E_{jlm}(r)\Y_{jlm}(\Omega)\,.
\label{E-VSH}
\ee
A similar expression is valid for the magnetic field, mapping $\H(\r)\to H_{jlm}(r)$\,.
As shown in \App{App:VSH}, Maxwell's equations (\ref{ME}) transform into a $6\times6$ matrix differential equation for the radial coordinate only, which in turn splits into two separate $3\times3$ blocks, one block corresponding to TE, the other to TM polarization. The TE block has the following form
\be
\left(\begin{array}{ccc}
 k\eps(r)&-\frac{1}{r}\frac{d}{dr} r&\frac{\alpha_l}{r}\\
 \frac{1}{r}\frac{d}{dr} r&k\mu(r)&0\\
 \frac{\alpha_l}{r}&0&k\mu(r)
\end{array}\right)
\left(\begin{array}{c}
E_{1lm}(r)\\
iH_{2lm}(r)\\
iH_{3lm}(r)
\end{array}\right)=0\,,
\label{TE-matrix}
\ee
with $E_{2lm}(r)=E_{3lm}(r)=H_{1lm}(r)=0$.
To obtain the corresponding matrix differential equation for TM polarization,
one needs to make the following exchange in \Eq{TE-matrix}:
\be
E_{jlm}(r)\leftrightarrow iH_{jlm}(r)\,, \ \ \ \eps(r)\leftrightarrow\mu(r)\,.
\label{exchange}
\ee
We therefore consider in the following only solutions for TE polarization, for generality of results keeping $\mu$ where appropriate  (even if $\mu=1$ everywhere, which is the case of non-magnetic systems).

Let us introduce new radial functions,
\be
\cE_j(r)=rE_{jlm}(r)\,,\ \ \
\cH_j(r)=riH_{jlm}(r)\,,
\label{EHrepl}
\ee
so that \Eq{TE-matrix} transforms into a simpler form
\be
\hM(k,r)\Ec(r) =0
\label{TE-matrix2}
\ee
with
\be
\hM(k,r)=\left(\begin{array}{ccc}
 k\eps(r)&-\frac{d}{dr} &\frac{\alpha}{r}\\
\frac{d}{dr}&k\mu(r)&0\\
 \frac{\alpha}{r}&0&k\mu(r)
\end{array}\right)
\label{Mkr}
\ee
and
\be
\Ec(r)=\left(\begin{array}{c}
\cE_1(r)\\
\cH_2(r)\\
\cH_3(r)
\end{array}\right)\,.
\label{Fdef}
\ee
Note that for brevity of notations we have also omitted here and almost everywhere below indices $l$ and $m$ (this includes  replacing $\alpha_l$ with just $\alpha$). Excluding $\cH_2$ and $\cH_3$, \Eq{TE-matrix2} transforms into the following  differential equation for $\cE_1$:
\be
\hL(k,r)\cE_1(r)=0\,,
\label{E1}
\ee
where $\hL(k,r)$ is a 2nd-order differential operator:
\be
\hL(k,r)=\mu(r)\frac{d}{dr}\frac{1}{\mu(r)}\frac{d}{dr} - \frac{\alpha^2}{r^2}+k^2\eps(r)\mu(r)\,.
\label{L}
\ee
Introducing a 1st-order vector differential operator
\be
\hO(k,r)=
\left(\begin{array}{c}
1\\
-\frac{1}{k\mu(r)}\frac{d}{dr}\\
-\frac{\alpha}{kr\mu(r)}
\end{array}\right)\,,
\label{O}
\ee
the full vectorial solution of \Eq{TE-matrix2} can then be written in the following compact form
\be
\Ec(r)=\hO(k,r)\cE_1(r)\,.
\label{Ec}
\ee

Equation~(\ref{GF-equ}) for the GF is transformed using the basis of the VSHs in a very similar way.
We first write the full $6\times 6$ GF more explicitly, in terms of four $3\times3$ blocks,
$$
\wG_k(\r,\r')=
\left(\begin{array}{cc}
\hG_k^{EE}&\hG_k^{EH}\\
\hG_k^{HE}&\hG_k^{HH}
\end{array}\right)\,,
$$
and then expand each block of the GF into the VSHs. The $EE$ block, for example, is expanded as
$$
\hG_k^{EE}(\r,\r')=\sum_{ij}\sum_{lm}[G^{EE}_{ij}(r,r')]_{lm} \Y_{ilm}(\Omega)\otimes\Y_{jlm}(\Omega')\,,
$$
where the single summation over $l,m$ is due to the spherical symmetry of the optical system. For the same reason, TE and TM parts of the GF separate from each other, with all the cross terms between different polarizations vanishing. Again, it is sufficient to find a general solution only for one of the two polarizations, then with the exchange \Eq{exchange} the solution in the other polarization takes exactly the same form. We therefore concentrate in the following on the TE block of the GF, which in the VSH basis has the form
$$
\frac{1}{rr'}\GF(r,r')\equiv \left(\begin{array}{ccc}
G_{11}^{EE}&G_{12}^{EH}&G_{13}^{EH}\\
G_{21}^{HE}&G_{22}^{HH}&G_{23}^{HH}\\
G_{31}^{HE}&G_{32}^{HH}&G_{33}^{HH}
\end{array}\right)
\,.
$$
Here, we have introduced for convenience, in full analogy with \Eq{EHrepl}, a new dyadic GF $\GF(r,r')$ which satisfies the following matrix differential equation:
\be
\hM(k,r)\GF(r,r') =\one\delta(r-r')\,,
\label{GFTE-equ}
\ee
where $\one$ is the $3\times3$ identity matrix.
It also follows from the general reciprocity relation \Eq{reciprocity} that
\be
\cG_{ij}(r',r)=\cG_{ji}(r,r')\,.
\label{rec}
\ee
in which $\cG_{ij}$  are the matrix elements of $\GF$.

\subsection{Dyadic Green's function for fixed $l$ and $m$}
\label{Sec:SGF}

First of all, we note that components $\cG_{12}$, $\cG_{21}$, $\cG_{23}$, and $\cG_{32}$ of the GF have discontinuities at $r=r'$ and component $\cG_{33}$ is irregular as it contains a $\delta$ function, as it immediately follows from \Eq{GFTE-equ} -- see also \App{App:GF} for details.
All other matrix elements of the GF, including the regular part of $\cG_{33}$ are continuous and finite for any finite $r$, $r'$, and complex $k$ (the same is true also for any component of the GF when $r\neq r'$).

It is important to note at this point that the slow convergence of the standard version of the RSE considered in \Sec{Sec:RSE} is actually caused by the presence of the $\delta$ function in  $\cG_{33}$ and by the fact that this $\delta$ function is expanded into static modes. Usually, expansions of $\delta$ functions into compete sets of regular smooth functions have very poor convergence. In the second version of the RSE presented in \Sec{Sec:RSE-elim}, this $\delta$ function is eliminated from the ML series. However, the slow convergence in that case is caused by two other $\delta$ functions added to elements $\cG_{11}(r,r')$ and $\cG_{22}(r,r')$, respectively. These $\delta$ functions are again represented by expansions, this time in terms of the RSs only, which makes this version of the RSE, from the point of its practical use, essentially similar to the first one.

The solution of \Eq{GFTE-equ} is derived in \App{App:GF}. As in the case of a homogeneous slab~\cite{Sam19}, the Green's dyadic can be written in the following compact way, using only the scalar function $\cG_{11}(r,r')$:
\be
\GF(r,r')=\hO(k,r)\otimes \hO(k,r') \cG_{11}(r,r') +\frac{\delta(r-r')}{k\mu(r)} (\one_2+\one_3)
\label{GF-sol}
\ee
with operator $\hO$ defined by \Eq{O} and $(\one_j)_{ii'}=\delta_{ii'}\delta_{ij}$. Element $\cG_{11}(r,r')$ of the dyadic GF satisfies the outgoing wave boundary conditions (for real $k$) and the following ordinary differential equation with a source
\be
\hL(k,r)\cG_{11}(r,r')=k\mu(r)\delta(r-r')\,,
\label{G11-equ}
\ee
where the operator $\hL$ is defined by \Eq{L}.

Equation (\ref{G11-equ}) can be easily solved for any spherically symmetric system, analytically (as done in \Sec{Sec:Sphere}) or numerically. The term $\one_3 \delta(r-r')$ in \Eq{GF-sol} is the singular part of $\cG_{33}(r,r')$ discussed above, which a true physical singularity of the dyadic GF. There is however an additional singular term $\one_2 \delta(r-r')$ which appears in \Eq{GF-sol} in order to compensate on a singularity emerging from second derivative which is appears after applying twice the operator $\hO(k,r)$ -- for more details, see \App{App:GF}.

Equation (\ref{G11-equ}) has the following explicit solution:
\be
\cG_{11}(r,r')=\frac{\cE_L(r_<)\cE_R(r_>)}{W},
\label{G11analyt}
\ee
where $r_<=\min (r,r')$, $r_>=\max (r,r')$, $\cE_{L(R)}(r)$ is the so-called left (right) solution, and
\be
W=\frac{\cE_L(r)\cE'_R(r)-\cE'_L(r)\cE_R(r)}{k\mu(r)}
\label{Wronskian}
\ee
is the Wronskian, which is independent of $r$. $\cE_{L(R)}(r)$ satisfies the corresponding homogeneous equation (\ref{E1}) and the left (right) boundary condition for the GF:
\be
\begin{array}{ll}
\cE_L(r)\propto r^{l+1}&\ {\rm at}\ r\to0\,,\\
\cE_R(r)\propto r h^{(1)}_l(kr)&\ {\rm at}\ r>R\,.
\label{ABC}
\end{array}
\ee
The first condition follows from the asymptotic behaviour of the operator \Eq{E1} at small $r$ and the regularity of the GF at the origin, while the second one is the outgoing boundary condition, assuming a constant  refractive index outside the system, e.g. $\eps(r)\mu(r)=1$. Here, $h^{(1)}_l(z)$ is the spherical Hankel function of 1st kind. Introducing the corresponding vector functions,
\be
\Ec_{L(R)}(r)=\hO(k,r)\cE_{L(R)}(r)
\label{FLR}
\ee
with $\hO$ given by \Eq{O}, the full dyadic GF takes the following form
\be
\GF(r,r')=\frac{\delta(r-r')}{k\mu(r)} \one_3+\frac{1}{W}\times\left\{
\begin{array}{ll}
\Ec_{L}(r)\otimes \Ec_{R}(r') & r<r'\\
\Ec_{R}(r)\otimes \Ec_{L}(r')  & r>r'\,,
\end{array}\right.
\label{GF-reg}
\ee
where the singular term $ \one_2 \delta(r-r')$, previously added to \Eq{GF-sol}, has now been removed, while the real, physical singularity of the dyadic GF remains. It is represented by the 1st term in \Eq{GF-reg}, clearly contributing to the static, $k=0$ pole of the GF. The 2nd term in \Eq{GF-reg} contains no spatial singularities, but it also brings in a significant contribution to the static pole of the GF, as we show in \Sec{Sec:StaticPole} below.

\subsection{Static pole of the dyadic GF}
\label{Sec:StaticPole}

To study the behaviour of the dyadic GF in the static limit and to find its residue at the $k=0$ pole, we introduce an auxiliary, $k$-independent matrix $\cR_{ij}$ defined in such a way that
\be
\GF(r,r')\to\frac{1}{k}\left(\begin{array}{ccc}
k^2\cR_{11}&k\cR_{12}&k\cR_{13}\\
k\cR_{21}&\cR_{22}&\cR_{23}\\
k\cR_{31}&\cR_{32}&\cR_{33}
\end{array}\right)
\label{GFzero}
\ee
at $k\to0$. Substituting \Eq{GFzero} into \Eq{GFTE-equ} and taking the limit $k\to0$, we find the following differential equation for matrix $\cR_{ij}$:
\be
\left(\begin{array}{ccc}
 0&-\frac{d}{dr} &\frac{\alpha}{r}\\
\frac{d}{dr}&\mu&0\\
 \frac{\alpha}{r}&0&\mu
\end{array}\right)
\left(\begin{array}{ccc}
\cR_{11}&\cR_{12}&\cR_{13}\\
\cR_{21}&\cR_{22}&\cR_{23}\\
\cR_{31}&\cR_{32}&\cR_{33}
\end{array}\right)
=\one\delta(r-r')\,,
\label{R-equ}
\ee
which is looking similar to \Eq{GFTE-equ}. Solving it in a similar way (see \App{App:GF} for details),  we find the residue of the dyadic GF at $k=0$:
\be
\hR(r,r') \equiv\left(\begin{array}{ccc}
0&0&0\\
0&\cR_{22}&\cR_{23}\\
0&\cR_{32}&\cR_{33}
\end{array}\right)
=-rr'\hnabla(r)\otimes\hnabla(r')g(r,r')
\label{R-sol1}
\ee
(note that matrices $\hR$ and $\cR_{ij}$ are not the same!).
In the VSH basis, the gradient operator has the form
\be
\hnabla(r)=
\left(\begin{array}{c}
0\\
\frac{\alpha}{r}\\
\frac{d}{dr}
\end{array}\right)\,,
\label{grad}
\ee
which is derived in \App{App:VSH}.  The new scalar GF $g(r,r')$ introduced in \Eq{R-sol1} satisfies the following equation
\be
\left[\frac{1}{r^2 \mu(r)}\frac{d}{dr}r^2 \mu(r)\frac{d}{dr} - \frac{\alpha^2}{r^2}\right]g(r,r')=\frac{\delta(r-r')}{r^2 \mu(r)}
\label{g-equ1}
\ee
and the boundary conditions that $g(r,r')$ is regular at $r,r\to0$ and vanishing at $r,r'\to\infty$. Element $\cR_{33}$  has a singularity equivalent to the first term in \Eq{GF-reg}. In the solution given by \Eq{R-sol1} this singularity is technically generated by the second mixed derivative of $g(r,r')$ -- see \App{App:GF} for details.

Interestingly, by varying the equation for the scalar GF, such as \Eq{g-equ1} for $g(r,r')$, the residue of the dyadic GF at the $k=0$ pole takes alternative representations, different from \Eq{R-sol1}, as discussed in more depths in  \Secs{Sec:ML}{Sec:pole} below and at the end of \App{App:Sphere}. Here we give one more representation, also derived in \App{App:GF}, which provides a natural link to the regular element $\cG_{11}$ of the dyadic GF in the limit $k\to0$:
\be
\hR(r,r') =
\hQ(r)
\otimes
\hQ(r')
 \tilde{g}(r,r') +
\frac{\delta(r-r')}{\mu(r)} (\one_2+\one_3)\,,
\label{R-sol2}
\ee
where we have introduced a new operator
\be
\hQ(r)=\lim_{k\to 0} k \hO(k,r)=-
\frac{1}{\mu(r)}
\left(\begin{array}{c}
0\\
\frac{d}{dr}\\
\frac{\alpha}{r}
\end{array}\right)
\label{Q}
\ee
and a new scalar GF $\tilde{g}(r,r,')$ satisfying an equation
\be
\left[\mu(r)\frac{d}{dr}\frac{1}{\mu(r)}\frac{d}{dr} - \frac{\alpha^2}{r^2}
\right]\tilde{g}(r,r')=\mu(r)\delta(r-r')
\label{g-equ2}
\ee
and the same boundary conditions as ${g}(r,r')$.
Since the operator in the square brackets in \Eq{g-equ2} is $\hL(0,r)$ [see \Eq{L}], we find
\be
\tilde{g}(r,r')=\lim_{k\to0} \frac{\cG_{11}(r,r')}{k}\,,
\label{gtilde}
\ee
in agreement with \Eq{G11-equ}. In fact,
the outgoing boundary condition for $\cG_{11}(r,r')$ transforms in the limit $k\to0$ into the vanishing boundary condition for $\tilde{g}(r,r')$ at $r,r'\to\infty$, owing to the asymptotic behaviour of the Hankel functions at a vanishing argument.
Similar to \Eq{GF-sol}, representation \Eq{R-sol2} of the static-pole residue of the GF introduces an additional explicit singularity $\one_2 \delta(r-r')/\mu(r)$ which is exactly compensated by the second mixed derivative in $\cR_{22}$.

\subsection{Static modes}
\label{Sec:static}

The scalar GF $g$ or $\tilde{g}$, defined by Eqs.\,(\ref{g-equ1}) or (\ref{g-equ2}), respectively, determines a complete set of static modes which can be used for expansion of the $k=0$ residue of the dyadic GF.
Note that with a replacement $r^2\mu(r)\to1/\mu(r)$, \Eq{g-equ1} transforms into \Eq{g-equ2}.
Let us therefore introduce a general second-order differential operator
\be
\hcL(r)=\frac{1}{w(r)}\frac{d}{dr} w(r)\frac{d}{dr} - \frac{\alpha^2}{r^2}\,,
\label{L-static}
\ee
where $w(r)$ is some weight function.
This operator generates an eigenvalue equation
\be
\left[\hcL(r)+ \lambda^2 \Theta(R-r)\right] \phi_\lambda(r)=0\,,
\label{psi}
\ee
where $\Theta(x)$ is the Heaviside step function.
The corresponding GF $G_\Lambda(r,r')$ satisfies an equation
\be
\left[\hcL(r)+ \Lambda \Theta(R-r)\right] G_\Lambda(r,r')=\frac{\delta(r-r')}{w(r)}\,.
\label{GLam}
\ee
Here, both $\phi_\lambda$ and $G_\Lambda$ obey vanishing boundary conditions at $r,r'\to \infty$ and regularity at the origin.
Note that $\lambda$ in \Eq{psi} is the eigenvalue, while $\Lambda$ in \Eq{GLam} is a parameter which can take any value.
Multiplying \Eq{psi} with $\phi_{\lambda'} w$, integrating the result over the full space, and then subtracting from it the same equation with $\lambda$ and $\lambda'$ interchanged, we obtain an orthogonality relation
$$
(\lambda^2-\lambda'^2)\int_0^R \ \phi_\lambda(r)\ \phi_{\lambda'}(r)w(r) dr =0\,.
$$

Then using the completeness of the set of functions $\phi_\lambda(r)$ and the symmetry of the GF, $G_\Lambda(r,r')=G_\Lambda(r',r)$, we obtain the following spectral representation
\be
G_\Lambda(r,r')=\sum_\lambda\frac{\phi_\lambda(r)\phi_\lambda(r')}{\Lambda-\lambda^2}\,,
\ee
valid within the system, i.e. for $r\leqslant R$.
Substituting it into \Eq{GLam} and using \Eq{psi}, we obtain a closure relation
$$
w(r)\Theta(R-r)\sum_\lambda{\phi_\lambda(r)\phi_\lambda(r')} = \delta(r-r')\,,
$$
confirming the completeness of the basis $\{\phi_\lambda\}$ within the system, and a normalization condition
\be
\int_0^R \ \phi_\lambda(r)\ \phi_{\lambda'}(r)w(r) dr = \delta_{\lambda\lambda'}\,,
\label{psi-norm}
\ee
which is combined here with the already proven orthogonality.

The scalar GF $g(r,r')$ contributing to the static pole of the dyadic GF via \Eq{R-sol1} is then given by a static-mode expansion
\be
g(r,r')=G_0(r,r')=-\sum_\lambda\frac{\phi_\lambda(r)\phi_\lambda(r')}{\lambda^2}
\label{G0}
\ee
with the static-mode basis $\{\phi_\lambda\}$  generated by \Eqs{L-static}{psi} with $w(r)=r^2\mu(r)$. In the case of a homogeneous sphere in vacuum, this basis, called volume-charge (VC) static-mode basis, was introduced in \cite{LobanovPRA19} and applied there successfully for treating both spherical and non-spherical systems.

\subsection{Resonant states and their normalization}
\label{Sec:Norm}

The wave function of RS $n$ is given by \Eq{Ec} with $k=k_n$ and $\Ec(r)=\Ec_n(r)$. The complex eigen wave number $k_n$ and  the first component of the vectorial wave function $\cE_1$  are solutions of the wave equation (\ref{E1}) with outgoing boundary conditions. From the general normalization of the RSs, \Eq{Norm}, we find, using the properties of the VSHs \Eqs{VSHortho}{VSH31} and integration by parts, the RS normalization:
\bea
1&=&\int_0^R (\eps \cE_1^2+\mu \cH_2^2+\mu\cH_3^2)dr+\frac{R}{k_n}\left.\left(\cH_2\cE_1'-\cE_1\cH_2'\right)\right|_{r=R_+}
\nonumber\\
&=&2\int_0^R \eps \cE_1^2dr+\frac{1}{k_n}\left[\left(\cE_1\frac{r}{\mu(r)}\cE_1'\right)'-\frac{2r}{\mu(r)}(\cE_1')^2\right]_{r=R_+}
\label{norm-rad}
\eea
where the prime means $d/dr$ and $R_+=R+0_+$ with a positive infinitesimal $0_+$.
Note that,  the second line in \Eq{norm-rad} presents exactly the same form of the RS normalization as was derived in \cite{MuljarovEPL10} for $\mu=1$, apart from the factor of 2 introduced later on in \cite{MuljarovOL18}.

\subsection{Mittag-Leffler series for the Green's dyadic}
\label{Sec:ML}

For its use in the RSE, the GF should have a dyadic product form. Such a product form is provided by applying the ML theorem~\cite{Arfken01}. Thanks to reciprocity, the RS poles of the GF contribute in a form of dyadic products of the corresponding RS fields $\Ec_n(r)$:
\bea
\GF(r,r')&=&
\sum_n\frac{\Ec_n(r)\otimes \Ec_n(r')}{k-k_n}
+ \frac{\hR(r,r')}{k}\,.
\label{MLgen}
 \eea
As for the static pole of the dyadic GF, its residue $\hR(r,r')$ introduced and studied in \Sec{Sec:StaticPole}
does not have a dyadic product form and therefore needs to be expanded into some basis states, which is done below.
In this section, we introduce and discuss three different ML representations of the dyadic GF. One more ML representation, with static-mode elimination, is provided in \Sec{Sec:Sphere} and illustrated in \Sec{Sec:Results}, in comparison with other versions.

\subsubsection*{1st ML representation}

Since the full dyadic GF $\GF(r,r')$ can be expressed in terms of its first element, as given by \Eq{GF-sol}, we concentrate here on finding a ML series for $\cG_{11}(r,r')$, a scalar GF satisfying \Eq{G11-equ} and outgoing boundary conditions. Equation~(\ref{G11-equ}) contains the same operator $\hL$, given by \Eq{L},  as appears in the wave equation (\ref{E1}) determining the electric field of the RSs in TE polarization:
\be
\hL(k_n,r)\cE_n(r)=0\,.
\label{E1RS}
\ee
Here we use for convenience index $n$ labelling the RSs, so that $\cE_1$ is replaced with $\cE_n$. Treating $\cG_{11}(r,r')$ as a function in the complex $k$-plane, we note that, thanks to \Eq{E1RS}, it has simple poles at $k=k_n$. Also, it vanishes as $1/k$ at large $k$, as it follows from \Eq{G11-equ}. Calculating the residues at the poles and then applying the ML theorem~\cite{Arfken01} to $\cG_{11}$, we find the following series representation:
\be\cG_{11}(r,r')=\sum_n \frac{\cE_n(r)\cE_n(r')}{k-k_n}\,,
\label{ML1}
\ee
where the field $\cE_n(r)$ is normalized according to \Eq{norm-rad}. The proof of \Eq{ML1} is very similar to that provided for non-magnetic systems in the Appendix of Ref.~\cite{MuljarovEPL10}; we therefore do not repeat it in this paper.

Taking into account the fact that the dyadic GF $\GF(r,r')$ has only simple poles at the RS wave numbers, $k=k_n$, and at $k=0$, as expressed by \Eq{MLgen}, we substitute the scalar ML expansion \Eq{ML1} into the general form of the dyadic GF, \Eq{GF-sol}. Comparing the result with \Eq{MLgen}, this leads to
\be
\Ec_n(r)=\hO(k_n,r)\cE_n(r)\,,
\label{Fn}
\ee
which is identical to \Eq{Ec} [the operator $\hO(k,r)$ is defined in \Eq{O}], provided that $\cE_1(r)$ in \Eq{Ec} is the RS field normalized according to \Eq{norm-rad}.

As for the $k=0$ pole, its residue is given by \Eq{R-sol1}, where the scalar GF $g(r,r')$ may be used in the form of the series \Eq{G0}. This results in the 1st ML representation of the dyadic GF:
\be
\GF(r,r')=
\sum_n\frac{\Ec_n(r)\otimes \Ec_n(r')}{k-k_n}
+\sum_\lambda \frac{\Ec_\lambda(r)\otimes \Ec_\lambda(r')}{k}\,,
\label{MLrep1}
 \ee
where the LM static-mode fields are given by
\be
\Ec_\lambda(r)=-r
\hnabla(r)\psi_\lambda(r)=-
\left(\begin{array}{c}
0\\
\alpha\psi_\lambda(r)\\
{r}\frac{d}{dr}\psi_\lambda(r)
\end{array}\right)\,,
\label{static1}
\ee
in accordance with \Eqs{LELM}{grad}. Here, $\psi_\lambda(r)=\phi_\lambda(r)/\lam$, and $\phi_\lambda(r)$ are the normalized eigen solutions of \Eq{psi} with $w(r)=r^2\mu(r)$.

The 1st ML representation given by \Eq{MLrep1} is identical to the general ML series~\Eq{GF-ML} introduced at the beginning of \Sec{Sec:RSsRSE}, which was also used for the conventional RSE in~\cite{DoostPRA14,LobanovPRA19}, though without any rigorous treatment of the static pole. Such a rigorous treatment and a proof of~\Eq{GF-ML} for spherically symmetric systems have now been provided above.

\subsubsection*{2nd ML representation}

It is also useful to apply the ML theorem to function  $\cG_{11}(r,r')/k$ which vanishes at $k\to \infty$ quicker than $\cG_{11}$  and takes a finite value at $k\to0$. In fact, $\cG_{11}$ is vanishing linearly in $k$ at $k\to0$ as can be seen from \Eq{G11-equ}. The ML series then takes the form
\be
\frac{1}{k}\cG_{11}(r,r')=\sum_n \frac{\cE_n(r)\cE_n(r')}{k_n(k-k_n)}\,.
\label{ML2}
\ee
Clearly, this series has a quicker convergence compared to its counterpart in \Eq{ML1}, due to the fact that $k_n\propto n$ at large $n$, which is a general property of Fabry-P\'erot modes in any optical system.

Substituting the series \Eq{ML2} for the GF into \Eq{G11-equ} and using \Eq{E1RS}, we obtain a closure relation,
\be
\eps(r)\sum_n \cE_n(r)\cE_n(r') = \delta(r-r')\,,
\label{closure-rad}
\ee
and a sum rule,
\be
\sum_n \frac{\cE_n(r)\cE_n(r')}{k_n} = 0\,,
\label{sum-rad}
\ee
which is equivalent to the fact that $\cG_{11}$ vanishes at $k=0$, as noted above -- see also \Eq{ML1}. Function $\cG_{11}/k$ is in turn finite and $\cG_{11}/k^2$ has a simple pole at $k=0$. Applying the ML theorem again, this time to $\cG_{11}/k^2$, we obtain
\be
\frac{1}{k^2}\cG_{11}(r,r')=\sum_n \frac{\cE_n(r)\cE_n(r')}{k^2_n(k-k_n)} -\frac{1}{k}\sum_n \frac{\cE_n(r)\cE_n(r')}{k^2_n} \,,
\label{ML3}
\ee
where the last term is noting else than $\tilde{g}(r,r')/k$, see \Eqs{gtilde}{ML2}.
The series representations given by \Eqsss{ML1}{ML2}{ML3} allow us to use the general solution \Eqs{GF-sol}{R-sol2}, for deriving a new ML series for the full dyadic GF. Using all three representations of $\cG_{11}(r,r')$, we first obtain
\bea
&&\hO(k,r)\otimes \hO(k,r') \cG_{11}(r,r')\nonumber\\
&&= \sum_n\frac{\hO(k_n,r)\cE_n(r)\otimes \hO(k_n,r')\cE_n(r')}{k-k_n}
\nonumber\\
&& -\frac{1}{k}\sum_n\frac{\hQ(r)\cE_n(r)\otimes \hQ(r')\cE_n(r')}{k^2_n}\,,
\label{OOG}
\eea
where the operators $\hO(k,r)$ and $\hQ(r)$ are given, respectively, by \Eqs{O}{Q}.
Note that the operator $\hQ(r)/k_n$ is the same as $\hO(k_n,r)$, apart from the first element which is vanishing in $\hQ(r)$. In the second, static-pole series in \Eq{OOG}, this operator can be upgraded to $\hO(k_n,r)$, by adding required terms to one diagonal and four off-diagonal elements of the dyadic GF. The terms added to the off-diagonal elements are however all vanishing, owing to the sum rule \Eq{sum-rad}, while the term added to the diagonal element $\cG_{11}$ can be converted into a $\delta$ function, thanks to the closure relation \Eq{closure-rad}. We therefore find a ML series for the dyadic GF in the following form:
\bea
\GF(r,r')&=& 
\sum_n\frac{\Ec_n(r)\otimes \Ec_n(r')}{k-k_n}
-\frac{1}{k}\sum_n{\Ec_n(r)\otimes \Ec_n(r')}\nonumber\\
 &&+\left[\frac{\one_1}{\eps(r)}+\frac{\one_2+\one_3}{\mu(r)}\right]\frac{\delta(r-r')}{k}\,,
\label{GF-ML2a}
\eea
where $\Ec_n(r)$ is given by \Eq{Fn}.

The 2nd ML representation given by \Eq{GF-ML2a} has no contribution of static modes and is equivalent to the general ML series \Eq{GF-ML2} introduced in \Sec{Sec:RSsRSE}. As it is shown in the example provided in \Sec{Sec:RSE-elim} above, the RSE based on this series has a rather slow convergence -- see also a comparison in \Sec{Sec:Results} below.

\subsubsection*{3rd ML representation}

In fact, the second series in \Eq{GF-ML2a} is very inefficient for representing elements $\cG_{11}$ and $\cG_{22}$ as it contains $\delta$ functions for both, expanded into sets of smooth functions. While the $\delta$ function in $\cG_{11}$ was added by hand, as described above, and thus can be easily removed, as done below, the static pole series for $\cG_{22}$ has a poor convergence due to the mixed second-order partial derivative, which also implicitly contains a $\delta$ function. To improve on this, we fist subtract in \Eq{GF-ML2a} the entire $k=0$ pole from $\cG_{22}$, which is given by
\be
\frac{\delta(r-r')}{k\mu(r)}\one_2 -\frac{1}{k\mu(r)\mu(r')}\sum_n \frac{\cE'_n(r)\cE'_n(r')}{k_n^2}\one_2 \,,
\label{sub}
\ee
with a singularity in the second term exactly compensating the $\delta$ function in the first one. We then add a regular representation of the $k=0$ pole of $\cG_{22}$, given by an expression
\be
-\frac{\alpha^2}{k}g(r,r')\one_2
\label{G22pole}
\ee
provided by the static-pole analysis of the GF, see \Eqs{R-sol1}{R22g}.

For the full GF to have a dyadic product form, we need to expand  \Eq{G22pole} into a complete set of functions. This can be any set which is complete within the system volume, $r\leqslant R$. The second-order differential operator \Eq{L-static} with $w(r)=r^2\mu(r)$ naturally generates such a basis, leading to \Eq{G0}.
Using this result, the full dyadic GF then takes the form:
\bea
\GF(r,r')&=&
\sum_n\frac{\Ec_n(r)\otimes \Ec_n(r')}{k-k_n}
+ \one_3\frac{\delta(r-r')}{k\mu(r)} \nonumber\\
&&-\frac{1}{k}\sum_n\frac{\hQ(r)\cE_n(r)\otimes \hQ(r')\cE_n(r')}{k_n^2}
\nonumber\\
&&+\frac{\one_2}{k}\sum_n\frac{\hQ_2(r)\cE_n(r) \hQ_2(r')\cE_n(r')}{k_n^2}
\nonumber\\
&&+\one_2\frac{\alpha^2}{k}\sum_\lambda \psi_\lambda(r)\psi_\lambda(r')\,,
\label{ML4}
 \eea
in which the second and the third series, when taken together, do not contain a singularity and are thus converging well, i.e. without an additional static-pole singularity error, in the same way as the first and the last series. Here, $\hQ_2(r)= -\mu^{-1}(r)d/dr$ is the second element of the vectorial operator $\hQ(r)$.
Equation (\ref{ML4}) is the 3rd ML representation provided in this paper. It contains an efficient summations over the RSs and static modes and thus should lead to a quicker version of the RSE, which is derived in \Sec{Sec:SRSE} below. The 3rd ML representation can be written in a more compact way by introducing general vectorial basis functions $\bpsi_j(r)$ representing the static pole:
\bea
\GF(r,r')&=&
\sum_n\frac{\Ec_n(r)\otimes \Ec_n(r')}{k-k_n}
+ \one_3\frac{\delta(r-r')}{k\mu(r)} \nonumber\\
&&+\frac{1}{k}\sum_j \bpsi_j(r)\otimes \bpsi_j(r')\,,
\label{ML4a}
 \eea
where index $j$ is running over all static modes ($\lambda$) once and  over all the RSs ($n$) twice, as it is clear from \Eq{ML4}.

We note that the 3rd ML representation given by \Eq{ML4} is not unique, and not only in the sense that different sets of static modes can be used, as mentioned above -- see also \cite{LobanovPRA19} where two different sets were used and \App{App:Sphere} in which three different sets of static mode are considered. In \Sec{Sec:Sphere} below we present one more ML representation of the Green's dyadic, having the same form as given by the more general \Eq{ML4a}. This 4th ML representation, suited for a homogeneous sphere, is focusing again on a complete elimination of static modes from the basis and developing a version of the RSE which is based on the RSs only.  Elimination of static modes is the main focus of this paper. Therefore, applying the RSE based on the 3rd ML representation \Eq{ML4} containing different sets of static modes will be done elsewhere.

\subsection{Resonant-state expansion}
\label{Sec:SRSE}

In the basis of the VSHs, Maxwell's equations (\ref{ME-pert}) for the perturbed system with spherical symmetry,
for TE polarization and $l$ and $m$ fixed, reduce to
\be
\left[\hat{\cal M}(k,r)+k\Delta\Pc(r)\right] \Ec(r) = 0\,,
\label{ME_pert}
\ee
where $\hat{\cal M}(k,r)$ is defined in \Eq{Mkr},
and
\be
\Delta\Pc(r)=
\left(\begin{array}{ccc}
 \Delta\eps(r)&0&0\\
 0&\Delta\mu(r)&0\\
 0&0&\Delta\mu(r)
\end{array}\right)
\label{DelP}
\ee
is the perturbation of the generalized permittivity within the sphere of radius $R$ containing the system.
Here $k$ is the eigen wave number, and \Eqsss{E-VSH}{EHrepl}{Fdef} define the components of the vector field $\Ec(r)$ of a perturbed RS.  The solution of \Eq{ME_pert} in terms of the dyadic GF  is given by
$$
\Ec(r)=-k\int_0^R\GF(r,r')\Delta\Pc(r')\Ec(r')dr'\,.
$$
 Using \Eq{ML4a}, we then find for the perturbed RS field:
$$
\Dc^{-1}(r)\Ec(r)=\sum_n a_n\Ec_n(r)+\sum_j b_j\bpsi_j(r)\,,
$$
where
$$
\Dc^{-1}(r)=\one+\frac{\Delta\mu(r)}{\mu(r)} \one_3\,.
$$
The expansion coefficients have the form
\bea
a_n&=&-\frac{k}{k-k_n}\int_0^R\Ec_n(r)\cdot \Delta\Pc(r)\Ec(r)dr \nonumber\\
&=&-\frac{k}{k-k_n}\left(\sum_{n'} V_{nn'} a_{n'} + \sum_{j'} V_{nj'} b_{j'} \right)\,,
\label{an}
\\
b_j&=&-\int_0^R\bpsi_j(r)\cdot\Delta\Pc(r)\Ec(r)dr \nonumber\\
&=&-\sum_{n'} V_{jn'} a_{n'} - \sum_{j'} V_{jj'} b_{j'}\,,
\label{bj}
\eea
where the matrix elements are given by
\be
\left(\begin{array}{cc}
 V_{nn'}&V_{nj'}\\
 V_{jn'}&V_{jj'}
\end{array}\right)
= \int_0^R dr  \left(\begin{array}{c}
\Ec_n\\
\bpsi_j
\end{array}\right)
\cdot \Delta\Pc \Dc
\left(\begin{array}{cc}
\Ec_{n'}& \bpsi_{j'}
\end{array}\right)
\label{Melem}
\ee
with
$$
\Delta\Pc\Dc=
 \Delta\eps(r)\one_1
 +\Delta\mu(r)\one_2
 +\frac{\mu(r)\Delta\mu(r)}{\mu(r)+\Delta\mu(r)}\one_3\,.
$$
Expressing the static amplitudes from \Eq{bj},
$$
b_j=-\sum_{j'}W_{jj'}\sum_n V_{j'n} a_n\,,
$$
\Eq{an} is transformed to the following matrix equation of the RSE:
\be
(k-k_n)a_n=-k\sum_{n'}\tilde{V}_{nn'} a_{n'}\,,
\label{RSE-gen}
\ee
where
$$
\tilde{V}_{nn'}={V}_{nn'}-\sum_{jj'} V_{nj} W_{jj'} V_{j'n'}\,,
$$
$W_{jj'}$ is the inverse of matrix $\delta_{jj'}+ V_{jj'}$, and $n$ labels all the basis RSs.
Again, \Eq{RSE-gen} can be symmetrized, as done at the end of  \Sec{Sec:RSE}.

For non-spherical perturbations which can mix states with different spherical numbers ($l,\,m$) and different polarizations, the formalism of the RSE and the key equation (\ref{RSE-gen}) remain essentially the same. The difference should appear in the matrix elements \Eq{Melem} which may be non-vanishing between TE and TM polarizations and between states with different pairs of ($l,\,m$) and ($l',\,m'$). Applying the RSE to such systems will be the subject of forthcoming publications.

\section{Application to a homogeneous sphere in vacuum}
\label{Sec:Sphere}

We now apply the formalism developed in \Sec{Sec:SSS} to a homogeneous sphere in vacuum, for its further use as the basis system in the RSE. The system is described by  uniform permittivity $\eps$ and permeability $\mu$ for $r\leqslant R$, where $R$ is the radius of the sphere, so that in the entire space
\bea
\eps(r)=1+(\eps-1)\Theta(R-r)\,,
\nonumber
\\
\mu(r)=1+(\mu-1)\Theta(R-r)\,.
\nonumber
\eea
It is useful to introduce at this point the refractive index $n_r$ and the impedance $\beta$ of the sphere defined as
\be
n_r=\sqrt{\eps\mu}\,,\ \ \ \
\beta=\sqrt{\eps/\mu}\,,
\label{nbe}
\ee
respectively, as both quantities contribute to the results obtained below.

Again, we concentrate in this section on TE polarization with fixed spherical quantum numbers $l$ and $m$. All results for TM polarization will then be exactly the same, provided that the replacement \Eq{exchange} is performed.

\subsection{Analytic form of the dyadic Green's function}
\label{Sec:GFsphere}

The analytic form of the dyadic GF is given by \Eqs{FLR}{GF-reg}, in terms of the left and right solutions, $\cE_{L,R}$. These have the following explicit form for the homogeneous sphere:
\be
\begin{array}{l}
\cE_L(r;k)=\left\{\begin{array}{ll}
J(n_rkr) & \ r\leqslant R\\
B_1J(kr)+B_2 H(kr) & \ r>R\,,
\end{array}
\right.
\smallskip
\\
\cE_R(r;k)=
\left\{\begin{array}{ll}
C J(n_rkr) + H(n_rkr) & r\leqslant R\\
B_3 H(kr) & r>R\,,
\end{array}
\right.
\end{array}
\label{ELR}
\ee
where $J(z)\equiv z j_l(z)$ and $H(z)\equiv z h_l^{(1)}(z)$, with $j_l(z)$ and $h_l^{(1)}(z)$ being, respectively, the spherical Bessel function and Hanken function of first kind. The $k$-dependent coefficients $C$, $B_{1}$, $B_{2}$, and $B_{3}$ in \Eq{ELR} are found by applying Maxwell's boundary conditions at $r=R$ and are provided in \App{App:Sphere}.

The Wronskian \Eq{Wronskian} contributing to the dyadic GF \Eq{GF-reg} is given by $W=i\beta$, see \Eq{W}, and the left and right vector functions
\Eq{FLR} have the following form inside the sphere ($r\leqslant R$):
\bea
\Ec_L(r;k)&=&\left(\begin{array}{c}
J(x)\\
-\beta J'(x)\\
-\alpha\beta J(x)/x
\end{array}\right)\,,
\label{FL}
\\
\Ec_R(r;k)&=&C(k)\Ec_L(r;k)+
\left(\begin{array}{c}
H(x)\\
-\beta H'(x)\\
-\alpha\beta H(x)/x
\end{array}\right)
\label{FR}
\eea
with $x=n_r k r$, $\alpha$ defined by \Eq{alpha}, $C(k)$ given by \Eq{Ccoef}, and primes meaning the derivatives of functions with respect to their arguments.

\subsection{Resonant states and their normalization}
\label{Sec:RSNorm}

The RS wave numbers $k_n$ are given by the poles of the coefficient $C(k)=N(k)/D(k)$, see \Eq{Ccoef}. Its denominator $D(k)$ thus determines the secular equation of the RSs in TE polarization:
\be
D(k_n)=\beta H(z)J'(n_r z)-H'(z)J(n_r z)=0\,,
\label{secular}
\ee
where $z=k_nR$. The RS wave functions which are given by \Eq{Fn} then take the form:
\be
\Ec_n(r)\equiv
\left(\begin{array}{c}
\cE_n(r)\\
\cK_n(r)\\
\cM_n(r)
\end{array}\right)
= A_n
\left(\begin{array}{c}
J(x)\\
-\beta J'(x)\\
-\alpha\beta J(x)/x
\end{array}\right)\,,
\label{EFn}
\ee
where $x=n_r k_n r$ ($r\leqslant R$) and $A_n$ are the normalization constants. The latter can be found by using the general normalization \Eq{norm-rad} or by calculating the residue at $k=k_n$ pole of the analytic GF:
\be
{A_n^2}=\frac{1}{i\beta}\lim_{k\to k_n} (k-k_n) C(k)\,,
\label{Andef}
\ee
see \Eq{GF-reg}.
Both ways are demonstrated in \App{App:Sphere}, leading to the same result: $A_n={\cal A}(n_r k_n R)$, where function ${\cal A}(z)$ is defined as
\be
\frac{1}{{\cal A}^2(z)R}= (\eps-1) J^2(z)+\eps(\mu-1)\left[\frac{\alpha^2}{z^2} J^2(z)+\frac{1}{\mu} J'^2(z)\right]\,.
\label{Az}
\ee

\subsection{Static pole expressed in terms of the RSs}
\label{Sec:pole}

We now find the explicit form of the static pole residue of the dyadic GF. It is given by general expressions \Eqs{R-sol1}{R-sol2}, in terms of
the scalar GFs $g$ and $\tilde{g}$ satisfying \Eqs{g-equ1}{g-equ2}, respectively. These GFs are provided in \App{App:Sphere} for the full space. Here we concentrate only on the region within the sphere, where they have the following form:
\be
\begin{array}{c}
g(r,r')=c_1\xi(r)\xi(r') + c_2\xi(r_<)\eta(r_>),
\smallskip
\\
\displaystyle
\frac{\tilde{g}(r,r')}{rr'}=\tilde{c}_1\xi(r)\xi(r') + \tilde{c}_2\xi(r_<)\eta(r_>)
\end{array}
\label{gg}
\ee
with
\be
\xi(r)=\left(\frac{r}{R}\right)^l\,,\ \ \ \ \eta(r)=\left(\frac{r}{R}\right)^{-l-1}\,,
\label{xi-eta}
\ee
and
\bea
{c}_1&=&-\frac{l+1}{l\mu^2}\, \tilde{c}_1= -\frac{1}{2l+1}\,\frac{1}{\mu R}\,\frac{(\mu-1)(l+1)}{\mu l+l+1}\,,
\nonumber \\
{c}_2&=&\frac{1}{\mu^2}\, \tilde{c}_2= -\frac{1 }{2l+1}\,\frac{1}{\mu R}\,.
\nonumber
\eea

We then find from \Eqs{R-sol1}{R-sol2} that the diagonal elements of the GF residue at the $k=0$ pole can be expressed in terms of the same functions $\eta(r)$ and $\xi(r)$:
\bea
\cR_{22}(r,r')&=&-\alpha^2 g(r,r')
\nonumber \\
&=&-\alpha^2 c_1\xi(r)\xi(r') -\alpha^2 c_2\xi(r_<)\eta(r_>)\,,
\nonumber \\
\cR_{33}(r,r')&=&\frac{\alpha^2\tilde{g}(r,r')}{\mu^2 rr'}
\nonumber \\
&=&-\alpha^2 \frac{l}{\mu(l+1)} c_1\xi(r)\xi(r') +\alpha^2 c_2\xi(r_<)\eta(r_>)\,,
\nonumber
\eea
which implies in particular that
\be
\cR_{22}(r,r')=-\cR_{33}(r,r')+c^2\xi(r)\xi(r')\,,
\label{R22R33}
\ee
where
\be
c^2=\frac{\alpha^2}{\mu R}\, \frac{\mu-1}{\mu l+l+ 1}
\label{c-def}
\ee
with $\alpha^2=l(l+1)$.

\subsubsection*{4th ML representation}

Now, instead of expressing the static pole of $\cG_{22}$, given by \Eq{G22pole} in terms of a complete set of static modes, as it is done in the 3rd ML representation of the dyadic GF, \Eq{ML4}, we use the link \Eq{R22R33} between the residues of the diagonal elements and the fact that $\cG_{33}$ has a quickly convergent expansion in terms of the RSs,
$$
-\cR_{33}(r,r')=\sum_n\frac{\alpha^2 \cE_n(r)\cE_n(r')}{\mu^2 k_n^2 rr'}= \sum_n \cM_n(r)\cM_n(r')\,,
$$
see \Eqsss{O}{ML4}{EFn}.

In fact, the above series has a quicker convergence, as compared to \Eq{ML1}, due to an additional power of $1/k_n$, see also \Fig{Fig:GF2} below illustrating it.
The last term in \Eq{R22R33} is looking like an effective single static mode added to the ML expansion, with a spatial profile $\xi(r)$ and a specific normalization given by the constant $c$.
We therefore arrive at one more ML expansion:
\bea
\!\!\!\!\!\!\!\!\GF(r,r')&=&
\sum_n\frac{\Ec_n(r)\otimes \Ec_n(r')}{k-k_n}
+ \one_3\frac{\delta(r-r')}{k\mu(r)} \nonumber\\
&&-\frac{1}{k}\sum_n\frac{\hQ(r)\cE_n(r)\otimes \hQ(r')\cE_n(r')}{k_n^2}
\nonumber\\
&&+\frac{\one_2}{k}\sum_n \cK_n(r)\cK_n(r')
\nonumber\\
&&
+\frac{\one_2}{k}\sum_n \cM_n(r)\cM_n(r')+\frac{\one_2}{k} c^2 \xi(r)\xi(r')
\label{ML5}
 \eea
with $\Ec_n(r)$ given by \Eq{EFn}, which defines its components $\cE_n(r)$, $\cK_n(r)$, and $\cM_n(r)$. Note that the difference between \Eqs{ML4}{ML5} is only in the last line: here, instead of using static modes, the same part of the GF is expressed in terms of the RS components $\cM_n(r)$.
The vectorial wave functions in the second line of these expansions can be conveniently expressed also in terms of the RS components:
$$
\frac{\hQ(r)\cE_n(r)}{k_n}=
\left(\begin{array}{c}
0\\
\cK_n(r)\\
\cM_n(r)
\end{array}\right)\,,
$$
using the definitions \Eqsss{O}{Q}{Fn}.

Equation~(\ref{ML5}) thus presents one more efficient ML representation of the dyadic GF provided in this paper, with a complete elimination of static modes from the basis,  and the static pole expressed in terms of the RS wave functions. The effective static mode, $\xi(r)$, present in the series can also be expressed in terms of the RS components, by using the fact that
$$
c\xi(r)=(\mu-1)\sqrt{\frac{l}{\mu}} \cK(0,r)= (\mu-1)\sqrt{\frac{l+1}{\mu}} \cM(0,r)\,,
$$
where
$$
\cK(0,r)=\lim_{k_n\to0} \cK_n(r)\,,\ \ \ \cM(0,r)=\lim_{k_n\to0} \cM_n(r)\,,
$$
see also functions $\cK(k,r)$ and $\cM(k,r)$ defined in \App{App:Sphere}.
A similar observation was made in~\cite{DoostPRA14} where a single static mode per each orbital quantum number $l$ was introduced. However, using the explicit form given by \Eqs{xi-eta}{c-def} might be more favorable for numerics.

The ML expansion \Eq{ML5} with static mode elimination is quickly convergent and is as efficient as the 3rd ML series \Eq{ML4} including static modes. It is thus well suited for its use in the RSE.
Furthermore, it suits the general form given by \Eq{ML4a}, provided that  functions $\bpsi_j(r)$
and indices $j$ are properly defined. The RSE formalism developed in \Sec{Sec:SRSE} can therefore be used in this case as well.

	\begin{figure}
\vspace{1.0mm}
\includegraphics*[clip,width=0.5\textwidth]{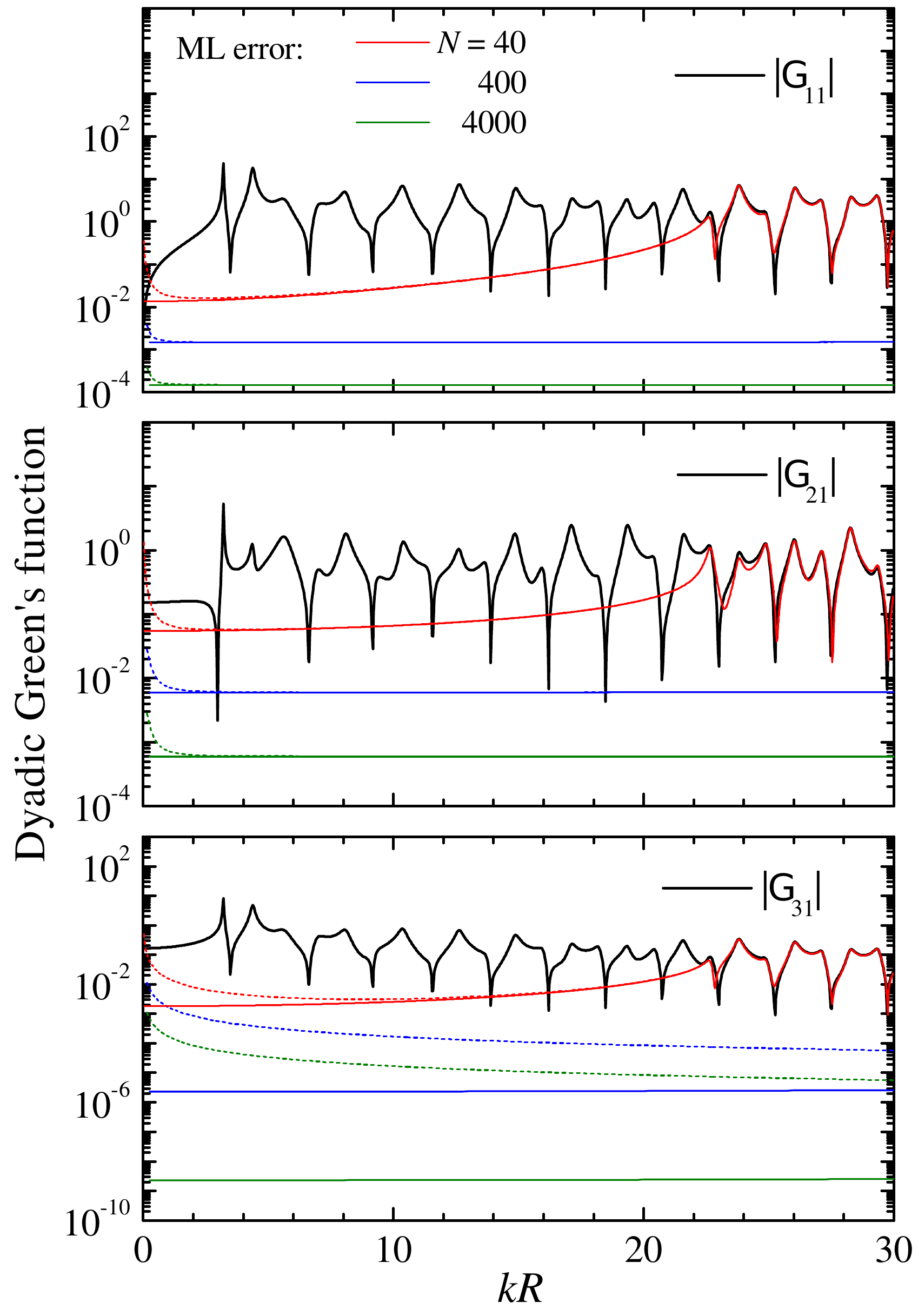}
		\caption{Elements $\cG_{11}$, $\cG_{21}$, and $\cG_{31}$ of the TE block of the dyadic GF (black solid lines) of a homogeneous sphere in vacuum and the absolute error for the 2nd (dashed lines) and 4th (solid lines) ML representations, given by \Eqs{GF-ML2a}{ML5}, respectively, for the number of RSs $N$ in the ML series as given. Results are shown as function of the real wave number $k$, for a magnetic sphere in vacuum, having $\eps=1$, $\mu=8$, and radius $R$. The coordinates of the GF are fixed at $r=0.5R$ and $r'=0.6R$.
 }
\label{Fig:GF1}
	\end{figure}
	\begin{figure}
\vspace{-1mm}
\includegraphics*[clip,width=0.5\textwidth]{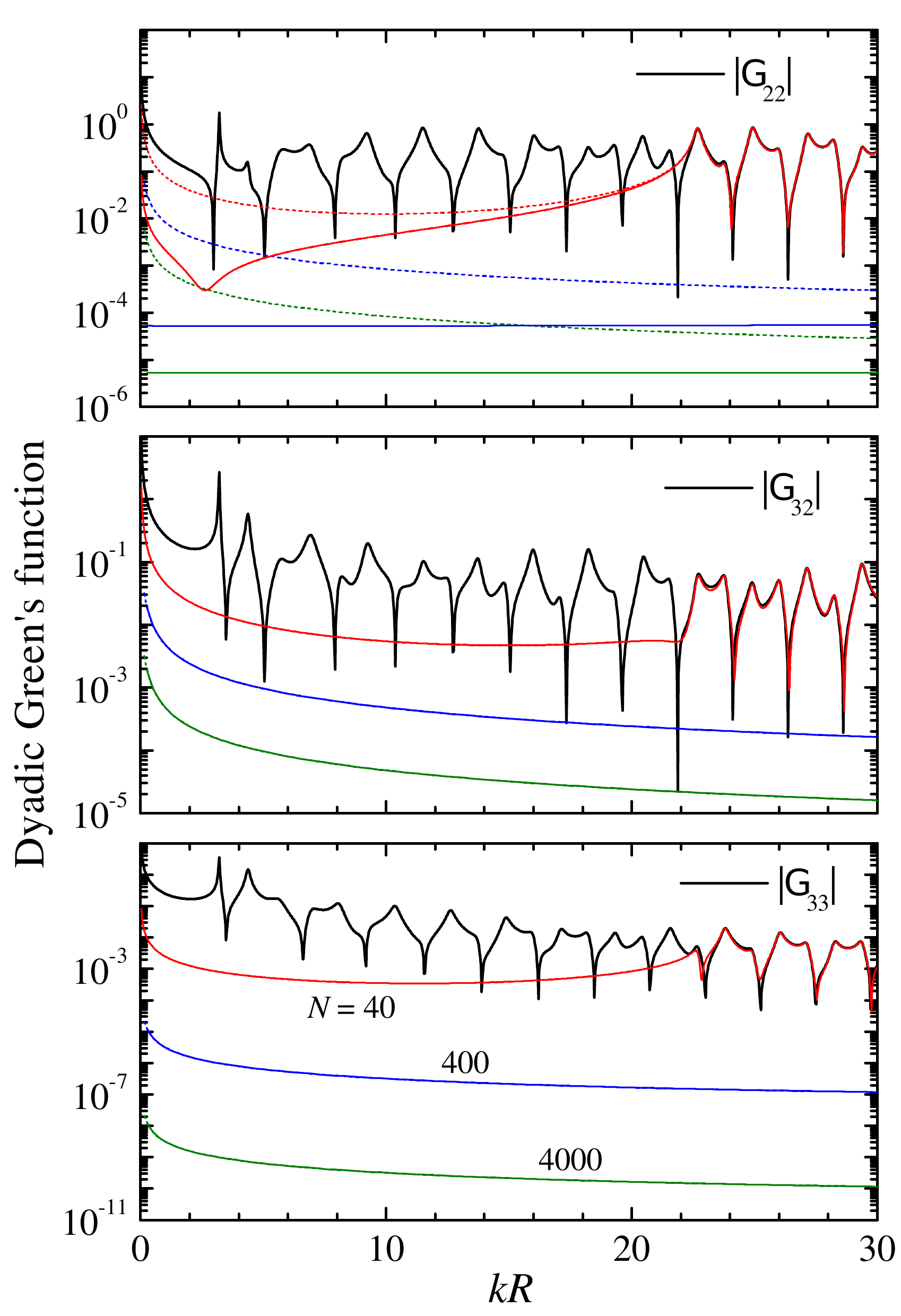}
		\caption{As \Fig{Fig:GF1} but for elements $\cG_{22}$, $\cG_{32}$, and $\cG_{33}$.
}
\label{Fig:GF2}
	\end{figure}

\section{Numerical results}
\label{Sec:Results}

In this section, we provide a few illustrations of the most important results obtained in this work. More illustrations of these results and application of the RSE to different systems, analyzing in particular the efficiency of the versions   introduced, will be presented elsewhere. Since the main focus of this paper is a proper elimination of static modes from the RSE, we concentrate below on the ML representations and the version of the RSE not containing static modes.

\subsection{Convergence of ML representations}
\label{Sec:NumGF}

Two different ML series introduced above, namely the 2nd and the 4th ML representations, which do not contain static-mode contributions are given, respectively,  by \Eqs{GF-ML2a}{ML5}.
These equations describe the TE block of the GF which can be equally used for TM polarization by swapping $\eps\leftrightarrow \mu$, as discussed in detail at the beginning of \Sec{Sec:SSS}. In particular, element $\cG_{11}$ of the TE block is responsible for the electric field in TE polarization, while elements $\cG_{22}$, $\cG_{23}$, $\cG_{32}$, and $\cG_{33}$ effectively describe the electric field in TM polarization. To address the TM polarization of a dielectric sphere with $\eps=8$ and $\mu=1$ taken for illustration, we therefore consider instead the TE block of the dyadic GF for a magnetic sphere with $\eps=1$ and $\mu=8$.

Figures~\ref{Fig:GF1} and \ref{Fig:GF2} show six elements of the TE block of the dyadic GF as a function of the real wave number $k$, for fixed $r=0.5R$ and $r'=0.6R$. Three other elements, $\cG_{12}$, $\cG_{13}$, and $\cG_{23}$, which are not shown, are quite similar to, respectively, $\cG_{21}$, $\cG_{31}$, and $\cG_{32}$.  The exact dyadic GF used for these plots is given by \Eq{GF-reg} with the left and right vector functions, $\Ec_{L}(r)$ and $\Ec_{R}(r)$, having the explicit analytic form provided in \Eqs{FL}{FR}, respectively. The exact GF is then compared with two ML representations, \Eqs{GF-ML2a}{ML5}, with the absolute difference shown in \Figs{Fig:GF1}{Fig:GF2} for both representations, for different numbers $N$ of RSs included in the expansion, in order to see how the ML series converge to the exact values.

It is clear that the first three elements of the Green's dyadic, illustrated in \Fig{Fig:GF1}, are not diverging at $k=0$, in agreement with the analysis provided in \Sec{Sec:SSS}. Furthermore, element $\cG_{11}$ vanishes at $k=0$, in accordance with \Eq{sum-rad}. However, the 2nd ML representation \Eq{GF-ML2a} demonstrates some footprints of the $1/k$ pole in these elements. This feature comes from the expansion of a $\delta$ function into the RSs, which is included in this ML series. The $\delta$ function contributes with a pre-factor $1/k$ [see \Eq{GF-ML2a}], explaining  the observed $1/k$ dependence of the error. The 4th ML representation \Eq{ML5} instead does not show any $1/k$ features and converges to the exact solution as $1/N$ for elements $\cG_{11}$ and $\cG_{21}$, and as $1/N^3$ for element $\cG_{31}$, much quicker than the 2nd one.

We further compare in \Fig{Fig:GF2} the two ML series, \Eqs{GF-ML2a}{ML5}, for elements $\cG_{22}$, $\cG_{32}$, and $\cG_{33}$ of the Green's dyadic. Physically, these components contain a $k=0$ pole contribution due to the spatial inhomogeneity of the system, which is clearly seen in  \Fig{Fig:GF2} as $1/k$ divergence. Note that this is additional to the longitudinal $\delta$-like singularity of the Green's dyadic of the homogeneous space~\cite{LevineCPAM50}, which should not be seen at $r\neq r'$. Here, the difference between the two representations is only in the $\cG_{22}$ component, which is again due to the fact that the 2nd ML representation \Eq{GF-ML2a} contains an expansion of a $\delta$ function contributing with a pre-factor $1/k$. This  additional divergent contribution, making the series representation  inefficient, is entirely eliminated in the 4th ML representation \Eq{ML5}, as it is clear from the top panel of \Fig{Fig:GF2}. As discussed in detail in \Sec{Sec:ML} above, this is the most significant improvement of the ML series implemented in the 3rd and also in the 4th ML representations, which results in a quickly convergent RSE, as demonstrated in \Sec{Sec:Shell} below. The ML series converge as $1/N$ for $\cG_{22}$ and $\cG_{32}$ components and as $1/N^3$ for $\cG_{33}$.

\subsection{RSE for a shell perturbation of a homogeneous sphere}
\label{Sec:Shell}

Consider a general spherical shell perturbation of the generalized permittivity \Eq{DelP} in the following form:
$$
\Delta\Pc(r)=\Theta(R_2-r)\Theta(r-R_1)\left[ \Delta\eps \one_1+\Delta\mu (\one_2+ \one_3)\right]\,,
$$
where  $R_1<R_2\leqslant R$.
This includes as special cases:\break (i) a homogeneous perturbation of the permittivity and permeability over the full volume of the sphere ($R_1=0$, $R_2=R$), which we call {\em strength} perturbation;  and\break (ii) reducing the radius of the sphere without changing its permittivity and permeability ($R_1>0$, $R_2=R$,\break $\Delta\eps=1-\eps$,  $\Delta\mu=1-\mu$), which is {\em size} perturbation.

For a shell perturbation within a region $R_1<r<R_2$, the matrix elements between RSs $\Ec_n(r)$ and $\Ec_m(r)$ of TE polarization, contributing to the RSE equation (\ref{RSE-gen}) are given by
\bea
V_{nm}&=&\Delta\eps \int _{R_1}^{R_2} \cE_n(r)\cE_m(r)dr
+\Delta\mu \int _{R_1}^{R_2} \cK_n(r)\cK_m(r)dr
\nonumber\\
&&+\frac{\mu\Delta\mu}{\mu+\Delta\mu} \int _{R_1}^{R_2} \cM_n(r)\cM_m(r)dr\,.
\label{Vnm}
\eea
Other elements, between RS wave functions $\Ec_n(r)$ and functions $\bpsi_j(r)$ representing the $k=0$ pole of the dyadic GF, or between function $\bpsi_j(r)$ and $\bpsi_{j'}(r)$ [see \Eqs{ML4a}{ML5}], have a similar form, and all necessary integrals contributing to the matrix elements are provided in \App{App:Sphere}.

As noted above, we illustrate in this paper only the versions of the RSE with static modes eliminated.

	\begin{figure}
\vspace{-1.5mm}
\includegraphics*[clip,width=0.5\textwidth]{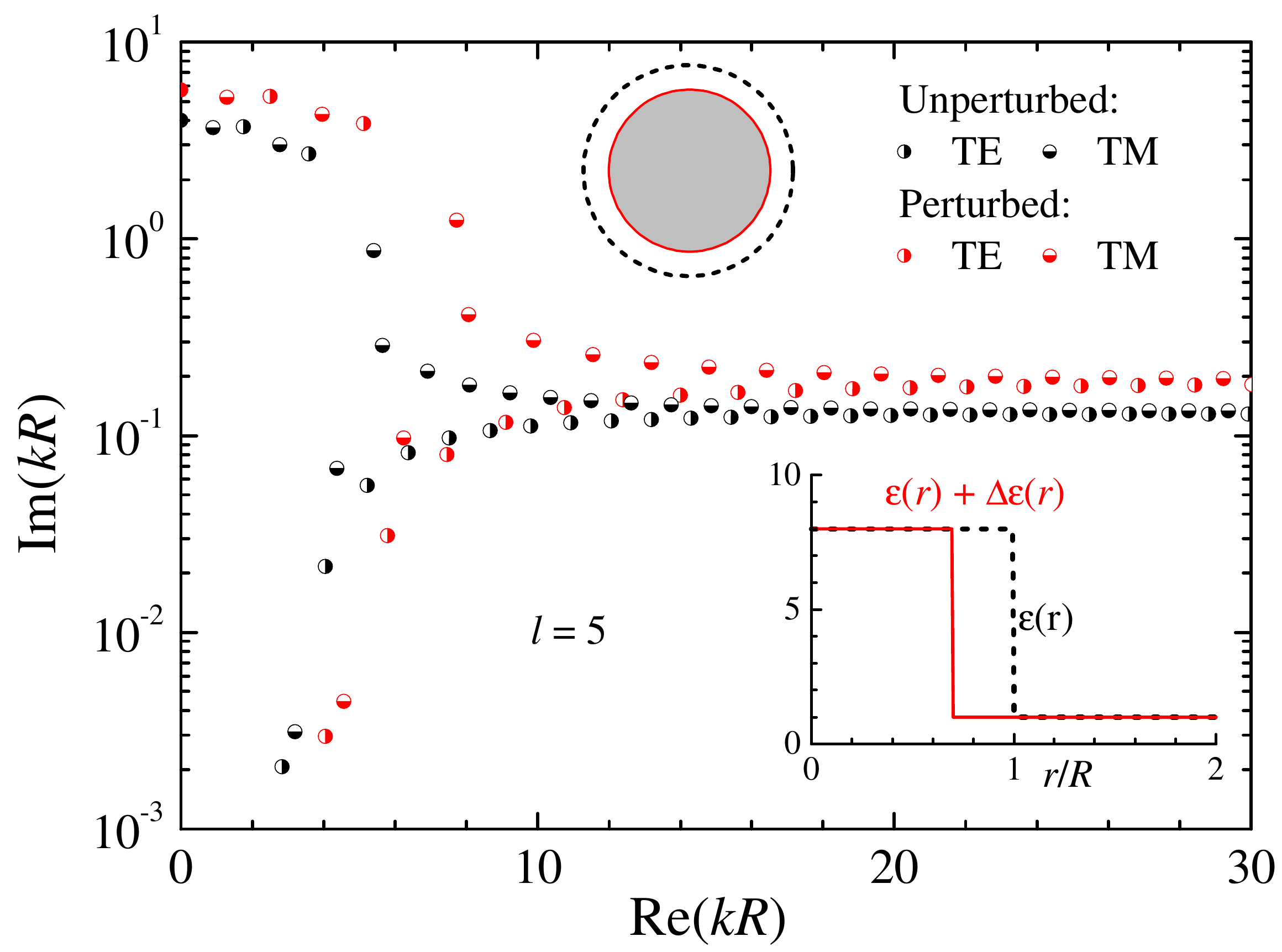}
\includegraphics*[clip,width=0.5\textwidth]{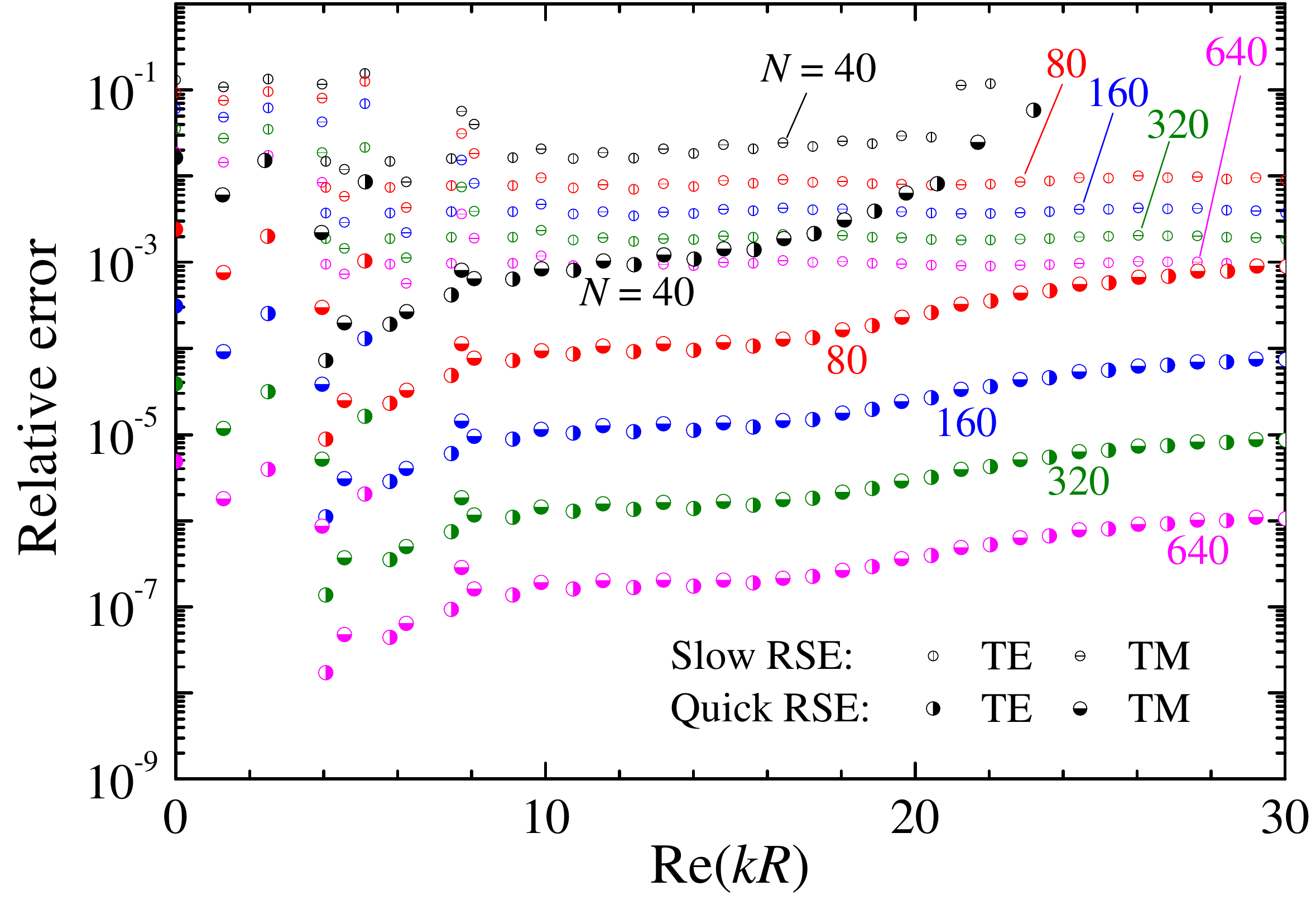}
		\caption{The RSE used for a size perturbation. Top: Wave numbers of TE and TM RSs for the unperturbed (black circles) and perturbed system (red circles). The unperturbed (perturbed) system is a homogeneous dielectric sphere in vacuum, with radius $R$ (0.7$R$), permittivity $\eps=8$, and permeability $\mu=1$.  The inset shows the profiles of the permittivity of the unperturbed and perturbed  systems (black dashed and red solid lines, respectively). Bottom: Relative error of the TE an TM RS wave numbers, calculated by the
RSE with static mode elimination. Results are shown for the slow RSE \Eq{RSE-mod3} and the quick RSE \Eq{RSE-gen}, corresponding, respectively to the 2nd and 4th ML representations, for different basis sizes $N$ as given.
}
\label{Fig:RSEsize}
	\end{figure}

	\begin{figure}
\vspace{1mm}
\includegraphics*[clip,width=0.5\textwidth]{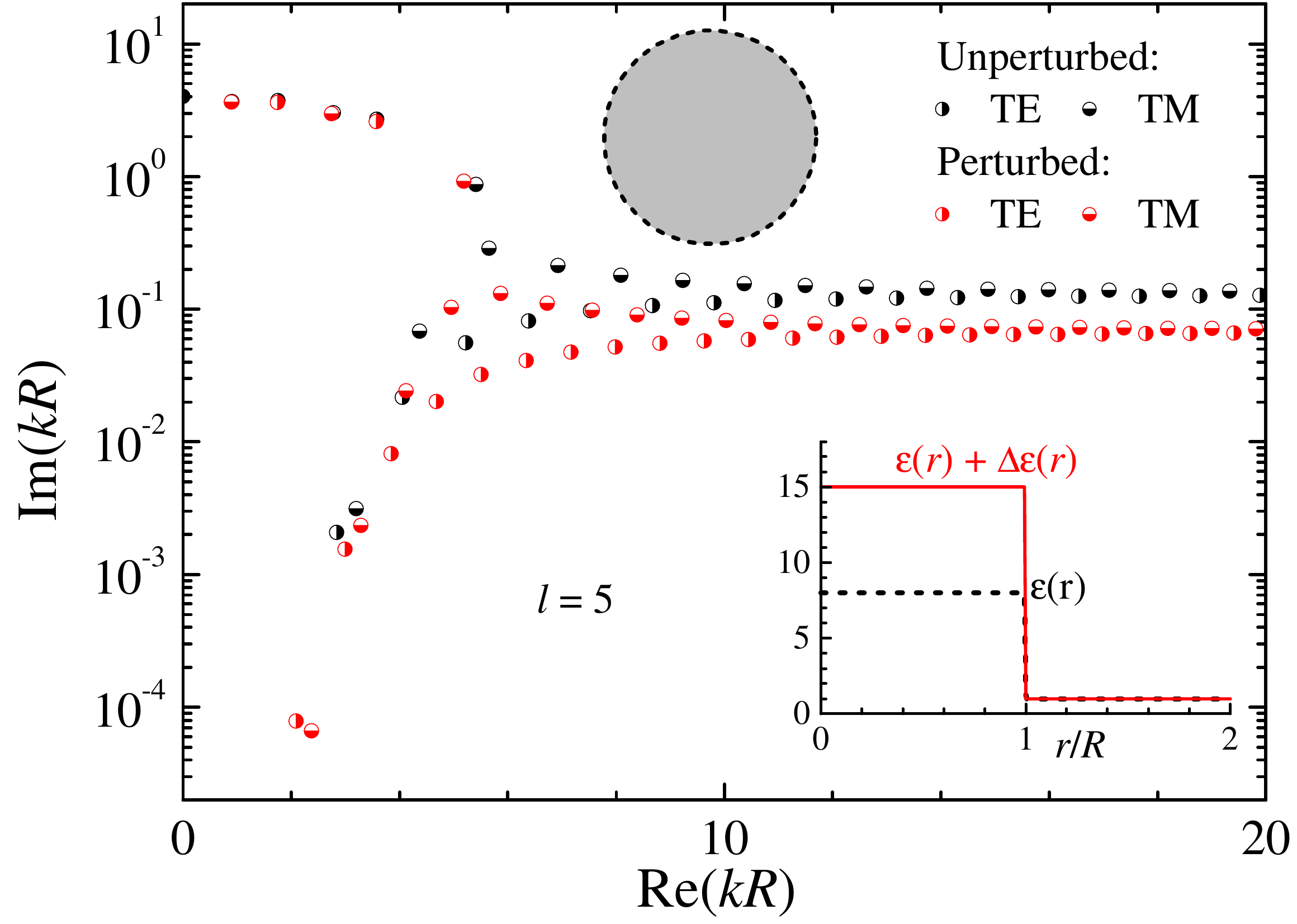}
\includegraphics*[clip,width=0.5\textwidth]{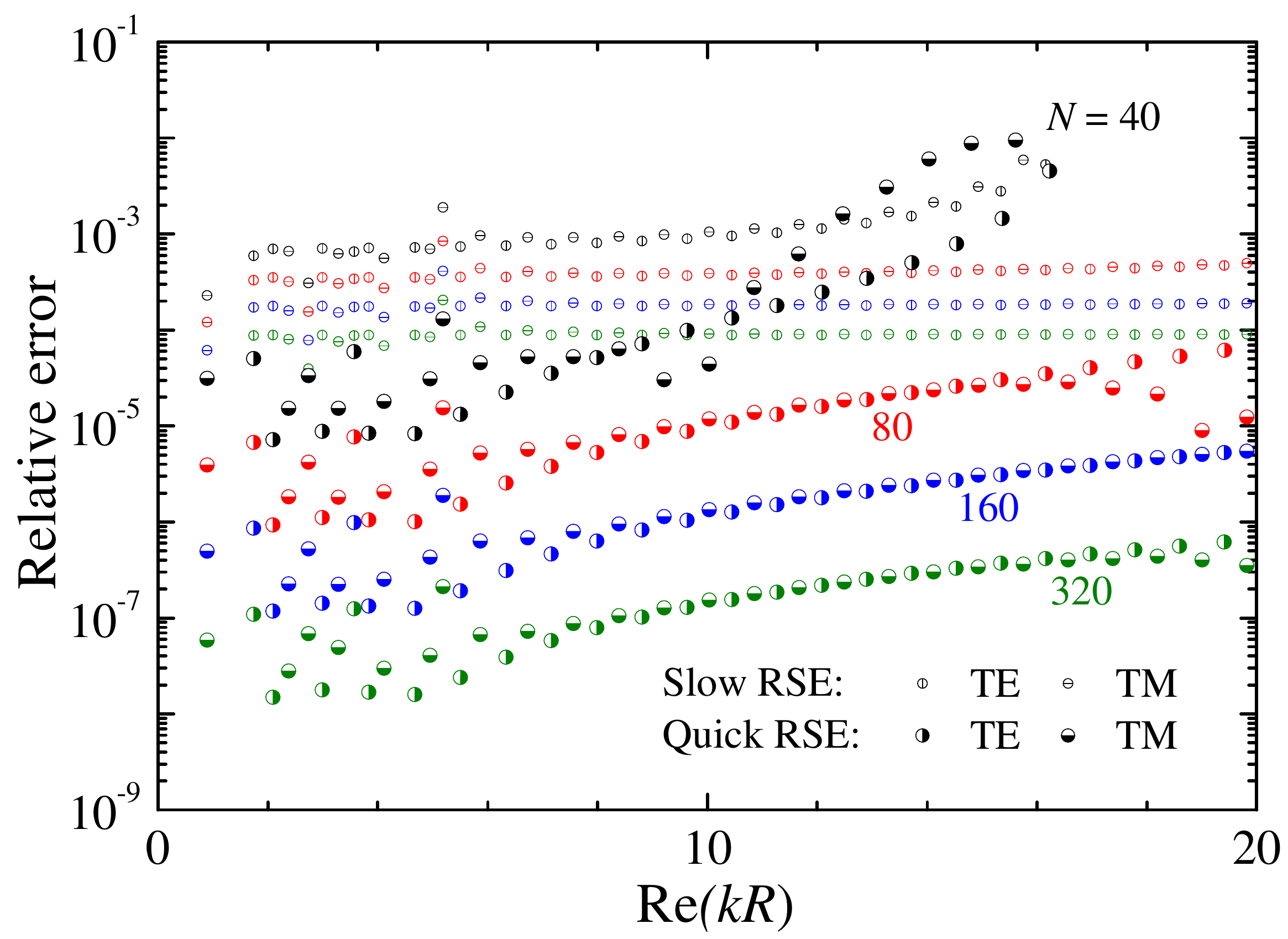}
		\caption{As \Fig{Fig:RSEsize} but for a strength perturbation of the sphere from $\eps=8$ to $\eps+\Delta\eps=15$.
}
\label{Fig:RSEstr}
	\end{figure}
\subsubsection{Size perturbation}

For the size perturbation, we modify the optical system from a dielectric sphere of radius $R$ and permittivity \mbox{$\eps=8$} to the same-permittivity sphere of radius $0.7R$.  We calculate both TE and TM modes of the smaller sphere using the slow and the quick versions of the RSE, both with static modes eliminated, and given by \Eqs{RSE-mod3}{RSE-gen}, respectively. These versions correspond, respectively, to the 2nd and 4th ML representations, given by \Eqs{GF-ML2a}{ML5}, which
we have illustrated in \Sec{Sec:NumGF} above, comparing with each other and with the analytic GF.

Figure~\ref{Fig:RSEsize} shows the unperturbed and perturbed RS wave numbers for both TE and TM polarizations and the relative error for the modes of the smaller sphere calculated via the slow and quick RSE, demonstrating the same level of efficiency for both polarizations.
Comparing the errors for different basis sizes $N$, it becomes clear that the quick (slow) RSE converges to the exact solution with relative error decreasing with $N$ as $1/N^3$ ($1/N$). Note that for this perturbation, the slow RSE has been already demonstrated for TM polarization in \Fig{Fig:slowRSE} above. Also note that for TE polarization, the quick RSE is identical to the original RSE formulated in \cite{MuljarovEPL10}. The latter was shown to  have a quick convergence to the exact solution in the TE polarization and is included as a special case in the generalized version of the RSE introduced in \Sec{Sec:SRSE} and illustrated in \Figs{Fig:RSEsize}{Fig:RSEstr}, which is a major fundamental result of this paper. This generalized version works equally well for both TE and TM polarizations, as demonstrated by \Figs{Fig:RSEsize}{Fig:RSEstr}, and is capable of treating, on the same level of efficiency, perturbations mixing TE and TM polarizations, as well as basis RSs with different spherical quantum numbers $l,m$.

\subsubsection{Strength perturbation}

We show for consistency the strength perturbation which is also very easy to verify, as this perturbation transforms an exactly solvable homogeneous sphere into another homogeneous sphere. Results are presented in \Fig{Fig:RSEstr}, showing that the convergence of both versions of the RSE is very similar to that in \Fig{Fig:RSEsize} for the size perturbation. Interestingly, for the strength perturbation, the overall level of errors is an order of magnitude smaller than for the size perturbation, even though the permittivity of the sphere in the strength perturbation is increased by almost a factor of two, while for the size perturbation the volume of the sphere is reduced by 2/3.

\section{Conclusions} 
\label{Sec:Con}

We have derived an analytic form of the electromagnetic Green's dyadic of an arbitrary spherically symmetric open optical system. Applying the formalism of vector spherical harmonics, the $6\times 6$ tensor of the dyadic Green's function (GF)  comprising the electric and magnetic field components on equal footing, is mapped onto an $(l,m)$-diagonal radially dependent tensor which further splits into two $3\times 3$ blocks separating transverse electric (TE) and transverse magnetic (TM) polarizations. In each polarization, the dyadic GF is expressed in terms of the so-called left and right solutions of a 2nd-order scalar differential equation determining its radial dependence. For a uniform distribution of the permittivity and permeability within a sphere, we have provided a fully explicit analytic solution for the dyadic GF, in terms of the spherical Bessel and Hankel functions.

We have studied analytically the pole structure of the dyadic GF, explicitly demonstrating for a general spherically symmetric system the link between the normalization of the  resonant states (RSs) and the pole residues of the dyadic GF at the RS frequencies. Using the analytic solution derived for the dyadic GF, we have also unambiguously determined its static pole residue, separating the regular part from the singularity described by a $\delta$ function and expanding this residue into different sets of static modes, as well as into the RSs themselves. This analysis has resulted in developing three different spectral representations of the dyadic GF of an arbitrary spherically symmetric system, which are called Mittag-Leffler (ML) representations. One more ML representation has also been found for a homogeneous sphere.

Different ML representations of the dyadic GF in turn generate different versions of the resonant-state expansion (RSE).
In this paper, we have formulated in total four different versions of the RSE, two of them having slow and the other two  quick convergence. Namely, they converge to the exact solution with the relative error proportional, respectively, to $1/N$ and $1/N^3$, where $N$ is the basis size used in the RSE. A comparative analysis of the four ML representations obtained in this work allowed us to reveal the source of poor convergence of the slow versions of the RSE, including the original one: Any expansion of the spatial singularity of the dyadic GF (related to its static pole) into a set of smooth functions, such as static modes or RS wave functions, slows down enormously the convergence of the RSE. With a  simple elimination of static modes as introduced at the beginning of this paper, the convergence of the RSE does not improve, remaining as slow as in the original version.

The paper presents a solution to this challenge, which is a proper removal of the singularity from the ML series for the dyadic GF. A detailed analysis of  the static pole of the GF allowed us to work out its regularized ML representations, with   $\delta$-function singularities separated from the series. This resulted in a new, quickly convergent version of the RSE, presented here in two variants -- with and without using static modes. While we have derived in this paper three different sets of static modes, also illustrating a significant freedom in their choice, we have focused in this work on the static-mode elimination. The main advantage of the RSE without static modes is that it depends only on a single parameter -- the number $N$ of the physical RSs of the basis system included, which is in turn determined by the truncation frequency in the complex plane.

We have illustrated the RSE with static-mode elimination on exactly solvable examples, used for verification and convergence study. These are perturbations of a homogeneous dielectric sphere in vacuum reducing its radius or uniformly changing its refractive index. Separating the static-pole singulary of the dyadic GF allowed us to accurately describe the effective charges induced by inhomogeneities of the permittivity and permeability, which manifest themselves in RS fields that are not divergence free. This is proven by demonstrating the same level of convergence of the RSE both with and without induced charges, realized in the selected examples, respectively, in TM and TE polarizations.

The developed generalization of the RSE, efficient in taking the induced charges into account, is the main fundamental result of this work. While illustrated here on spherical systems only, this generalized RSE is capable of treating, on the same level of efficiency, non-spherical perturbations mixing TE and TM polarizations and different spherical quantum numbers $(l,m)$, which will be the focus of follow-up publications.  Furthermore, as the RSE always maintains the completeness, it offers a unique tool for finding numerically exactly the full dyadic GF of an arbitrary non-spherical open optical system. Presently, this aim is not achievable by any other means.

\appendix
\section{Vector spherical harmonics: definitions, properties, and application}
\label{App:VSH}

The VSHs are defined by \Eq{VSH} with $Y_{lm}(\Omega)$ being the scalar spherical harmonics which are given by the following {\em real} functions:
\begin{equation}
Y_{lm}(\Omega) = \sqrt{\frac{2l+1}{2}\frac{(l-|m|)!}{(l+|m|)!}}P^{|m|}_l(\cos\theta)\chi_m(\varphi)\,,
\label{Eq:Y}
\end{equation}
where $P^m_l(x)$ are the associated Legendre polynomials, and
\be
\chi_m(\varphi)=\left\{
\begin{array}{lll}
\pi^{-1/2}\sin(m\varphi) & {\rm for} & m<0\, \\
(2\pi)^{-1/2} & {\rm for} & m=0\, \\
\pi^{-1/2}\cos(m\varphi) & {\rm for} & m>0\,.
\end{array}
\right. \label{chi-n}
\ee
The orthonormality condition for the VSHs has the form~\cite{LobanovPRA18}
\begin{equation}
\int \Y_{ilm}(\Omega)\cdot \Y_{i'l'm'}(\Omega)d \Omega=\delta_{ii'}\delta_{ll'}\delta_{mm'}\,,
\label{VSHortho}
\end{equation}
where $d\Omega=\sin\theta d\theta d\varphi$.
From the definition \Eq{VSH} and the orthogonality \Eq{VSHortho} follow useful properties of the VSHs:
\bea
\Y_{2lm}(\Omega)\times \Y_{1lm}(\Omega)\cdot {\bf e}_r &=& \Y_{1lm}^2(\Omega)\,,
\nonumber
\\
\Y_{3lm}(\Omega)\times \Y_{1lm}(\Omega)\cdot {\bf e}_r &=& 0\,
\label{VSH31}
\eea
(here ${\bf e}_r=\r/r$), which are helpful for deriving the RS normalization \Eq{norm-rad}.

Substituting the expansions \Eq{E-VSH} of $\E(\r)$ and $\H(\r)$ into \Eq{RS-equ}, we obtain for the first Maxwell's equation
\bea
0&=&k\eps(r)\E(\r)-\nabla\times i\H(\r)\nonumber\\
&=&k \eps(r)\sum_{jlm} E_{jlm}(r)\Y_{jlm}-\nabla\times \sum_{jlm} iH _{jlm}(r)\Y_{jlm}\nonumber\\
&=&\left.\sum_{lm}\right\{ \left[k\eps(r)E_{1lm}(r)-\frac{1}{r}\frac{d}{dr}r iH_{2lm}(r)-\frac{\alpha_l}{r}iH_{3lm}(r)\right]
\nonumber\\
&&\times\Y_{1lm}+\left[k\eps(r)E_{2lm}(r)+\frac{1}{r}\frac{d}{dr}r iH_{1lm}(r)\right]\Y_{2lm}
\nonumber\\
&&\left.+\sum_{lm} \left[k\eps(r)E_{3lm}(r)+\frac{\alpha_l}{r}iH_{1lm}(r)\right]\Y_{3lm}\right\}\,,
\nonumber
\eea
using $\nabla\times f(r)\Y_{jlm}=f(r)\nabla\times\Y_{jlm}+\r\times\Y_{jlm} f'(r)/r$ and results for $\nabla\times\Y_{jlm}$ and $\r\times\Y_{jlm}$  provided in~\cite{LobanovPRA18}. Deriving a similar expression for the second Maxwell's equation and using the orthonormality of the VSHs, \Eq{RS-equ} transforms into
\be
\left(\begin{array}{cccccc}
 k\eps&0&0&0&-\frac{1}{r}\frac{d}{dr} r&\frac{\alpha_l}{r}\\
 0&k\eps&0&\frac{1}{r}\frac{d}{dr} r&0&0\\
 0&0&k\eps&\frac{\alpha_l}{r}&0&0\\
 0&-\frac{1}{r}\frac{d}{dr} r&\frac{\alpha_l}{r}&k\mu&0&0\\
 \frac{1}{r}\frac{d}{dr} r&0&0&0&k\mu&0\\
 \frac{\alpha_l}{r}&0&0&0& 0&k\mu
\end{array}\right)
\left(\begin{array}{c}
E_{1lm}\\
E_{2lm}\\
E_{3lm}\\
iH_{1lm}\\
iH_{2lm}\\
iH_{3lm}
\end{array}\right)=0\,.
\label{full-matrix}
\ee
The matrix in \Eq{full-matrix} can be made block-diagonal, by simultaneous swapping of its columns and rows, so that the full $6\times6$ problem for each ($l,m$) splits into two blocks, one for TE, the other for TM polarization.

Let us also express the gradient operator in the basis of the VSHs. For an arbitrary scalar field $f(\r)$, we obtain
\bea
\nabla f(\r)&=&\nabla \sum_{lm} f_{lm}(r)Y_{lm}\nonumber\\
&=& \sum_{lm} [f_{lm}(r)\nabla Y_{lm}+Y_{lm}\nabla f_{lm}(r)]
\nonumber
\\
&=& \sum_{lm} \left[\Y_{2lm}\frac{\alpha_l}{r} f_{lm}(r)+ \Y_{3lm}\frac{d}{dr}f_{lm}(r)\right],\nonumber
\eea
using the definition of the VSHs, \Eq{VSH}. Then, for fixed $l$ and $m$, the gradient operator is given by
 \Eq{grad}.

\section{Derivation of the spherically symmetric dyadic GF}
\label{App:GF}

In this Appendix, we derive \Eqsss{GF-sol}{R-sol1}{R-sol2}, describing the analytic behaviour of the dyadic GF of a spherically symmetric open optical system and its residue at the static, $k=0$ pole.

First of all, it is straightforward to obtain \Eq{G11-equ}, by excluding $\cG_{21}$ and $\cG_{31}$ from the simultaneous equations given by \Eq{GFTE-equ}. We then find from the same equation that
$$
\cG_{21}(r,r')=-\frac{1}{k\mu(r)}\frac{d}{dr} \cG_{11}(r,r')
$$
and, using the reciprocity relation \Eq{rec}, obtain
$$
\cG_{12}(r,r')=\cG_{21}(r',r)=-\frac{1}{k\mu(r')}\frac{d}{dr'} \cG_{11}(r,r')\,.
$$
From the last equation and again, from  \Eq{GFTE-equ}, we then find
\bea
\!\!\!\!\!\!\!\!\!\!\!\cG_{22}(r,r')&=&\frac{\delta(r-r')}{k\mu(r)}-\frac{1}{k\mu(r)} \frac{d}{dr} \cG_{11}(r,r')
\nonumber\\
&=&\frac{\delta(r-r')}{k\mu(r)}+\frac{1}{k^2\mu(r)\mu(r')} \frac{d}{dr} \frac{d}{dr'}\cG_{11}(r,r').
\label{G22}
\eea
Note that $\cG_{22}$ is a regular component of the dyadic GF, and the $\delta$ function which appears explicitly in \Eq{G22} is needed to exactly compensate on the same singularity of the second term in \Eq{G22}, which is due to the double differentiation. In fact, integrating \Eq{G11-equ}, we find
\be
\frac{1}{k\mu(r)}\frac{d}{dr} \cG_{11}(r,r')= f(r,r')+\Theta(r-r')\,,
\label{kmuG}
\ee
where $f(r,r')$ is a continuous regular function and $\Theta(x)$ is the Heaviside step function. Then
$$
\frac{1}{k\mu(r)}\frac{d}{dr} \frac{d}{dr'} \cG_{11}(r,r') = \frac{d}{dr'} f(r,r')-\delta(r-r')\,,
$$
demonstrating the above mentioned singularity.

We next evaluate from \Eq{GFTE-equ}
\bea
\cG_{32}(r,r')&=&\cG_{23}(r',r)=-\frac{\alpha}{kr\mu(r)}\cG_{12}(r,r')\nonumber\\
&=&\frac{\alpha}{k^2r\mu(r)\mu(r')}\frac{d}{dr'} \cG_{11}(r,r')
\nonumber
\eea
and
\be
\cG_{31}(r,r')=\cG_{13}(r',r)=-\frac{\alpha}{kr\mu(r)}\cG_{11}(r,r')\,.
\label{G31}
\ee
The last element of the GF may be evaluated by combining \Eqs{GFTE-equ}{G31}:
\bea
\cG_{33}(r,r')&=&\frac{\delta(r-r')}{k\mu(r)}-\frac{\alpha}{kr\mu(r)} \cG_{13}(r,r')
\nonumber
\\
&=&\frac{\delta(r-r')}{k\mu(r)}+\frac{\alpha^2}{k^2r\mu(r)r'\mu(r')} \cG_{11}(r,r')\,.\nonumber
\eea
Clearly this element is irregular as it contains a singular term which is not compensated by any derivative.
Collecting all the elements of the dyadic GF derived above, we arrive at \Eq{GF-sol}.

Looking at the elements $\cG_{12}$, $\cG_{21}$, $\cG_{23}$, and $\cG_{32}$, evaluated above, we see that all of them have discontinuities at $r=r'$, as they are expressed in terms of the first derivative of  $\cG_{11}$, which  is discontinuous at $r=r'$, according to \Eq{kmuG}.

Now we derive in a similar way the two forms of the solution of \Eq{R-equ}, provided in \Sec{Sec:StaticPole}, which are \Eqs{R-sol1}{R-sol2}. Excluding $\cR_{12}$ and $\cR_{32}$ from \Eq{R-equ}, we obtain a differential equation for $\cR_{22}$:
$$
\left[-\frac{1}{\alpha^2}\frac{d}{dr}r^2 \mu(r)\frac{d}{dr} + \mu(r) \right]\cR_{22}(r,r')=\delta(r-r')
$$
which becomes \Eq{g-equ1} after a substitution
\be
\cR_{22}(r,r')=-\alpha^2 g(r,r')\,.
\label{R22g}
\ee
Other elements can be found straightforwardly from \Eq{R-equ}:
$$
\cR_{32}(r,r')=\cR_{23}(r',r)=\frac{r}{\alpha}\frac{d}{dr} \cR_{22}(r,r')\,.
$$
and
$$
\cR_{33}(r,r')=\frac{r}{\alpha}\frac{d}{dr} \cR_{23}(r,r')=\frac{rr'}{\alpha^2}\frac{d}{dr} \frac{d}{dr'} \cR_{22}(r,r')
\,.
$$
Then, using the link \Eq{R22g}, we obtain the solution \Eq{R-sol1}.

On the other hand, one can use instead element $\cR_{33}$ as a start point. Introducing its regular part $\cR_{33}^R$,
$$
\cR_{33}(r,r')=\cR_{33}^R(r,r')+\frac{\delta(r-r')}{\mu(r)}\,,
$$
we obtain from \Eq{R-equ}
$$
\cR_{23}(r,r')=\cR_{32}(r,r')=\frac{1}{\mu(r)}\frac{d}{dr} \frac{r\mu(r)}{\alpha}\cR_{33}^R(r,r')
$$
and the following differential equation for $\cR_{33}^R$:
$$
\left[-\frac{d}{dr}\frac{1}{\mu(r)}\frac{d}{dr}\frac{r\mu(r)}{\alpha} + \frac{\alpha}{r} \right]\cR_{33}^R(r,r')=-\frac{\alpha}{r\mu(r)}{\delta(r-r')}{}\,.
$$
Element $\cR_{22}$ can then be found, by noting that
$$
\cR_{12}(r,r')=-\frac{r\mu(r)}{\alpha} \cR_{32}(r,r')\,,
$$
so that, again, from \Eq{R-equ} we obtain
\bea
&&\cR_{22}(r,r')=\frac{\delta(r-r')}{\mu(r)}-\frac{1}{\mu(r)}\frac{d}{dr}\cR_{12}(r,r')
\nonumber
\\
&&=\frac{\delta(r-r')}{\mu(r)}+\frac{1}{\mu(r)\mu(r')}\frac{d}{dr}\frac{d}{dr'}\frac{r\mu(r)r'\mu(r')}{\alpha^2} \cR_{33}^R(r,r')\,.\nonumber
\eea
Introducing a scalar GF $\tilde{g}(r,r')$ such that
$$
r\mu(r)r'\mu(r') \cR_{33}^R(r,r')=\alpha^2 \tilde{g}(r,r')\,,
$$
we arrive at \Eqs{R-sol2}{g-equ2}.

\section{Homogeneous sphere in vacuum}
\label{App:Sphere}

\subsubsection{Green's function $\cG_{11}$}
\label{App:C1}

The general form of the scalar GF $\cG_{11}(r,r')$ is given by \Eq{G11analyt}.
For a homogeneous sphere in vacuum, the wave equation (\ref{E1}) with the operator $\hL$ given by \Eq{L} becomes
\bea
\left(\frac{d^2}{dr^2} -\frac{\alpha^2}{r^2}+n_r^2k^2\right)\cE(r)=0 &\ \ \ r\leqslant R\,,
\label{WE1}
\\
\left(\frac{d^2}{dr^2} -\frac{\alpha^2}{r^2}+k^2\right)\cE(r)=0 &\ \ \ \  r>R\,,
\label{WE2}
\eea
which are both wave equations for a homogeneous space in 3D. Their solution can therefore be expressed in terms of
spherical Bessel functions:
\be
\cE(r)=\left\{\begin{array}{ll}
C_1 J(n_rkr)+C_2H(n_rkr) & r\leqslant R\\
B_1J(kr)+B_2 H(kr) & r>R\,,
\end{array}
\right.
\label{Egeneral}
\ee
see \Sec{Sec:GFsphere} for the definition of $J(z)$ and $H(z)$. Note that \Eqs{WE1}{WE2}  have to be solved together with the boundary conditions of continuity of $\cE(r)$ and $\frac{1}{\mu(r)}\frac{d}{dr}\cE(r)$, following from \Eqs{E1}{L}. These boundary conditions are equivalent to Maxwell's boundary conditions of the continuity of the tangent components of the electric and magnetic fields, as it is clear from \Eqsss{Fdef}{O}{Ec}.  The coefficients in \Eq{Egeneral} are thus found from these boundary conditions and the additional ``left'' and ``right'' boundary conditions \Eq{ABC}. The latter lead to $C_2=0$ in the left and $B_1=0$ in the right solution. Also, without loss of generality, we have chosen $C_1=1$ in the left and $C_2=1$ in the right solution. The left and right solutions $\cE_{L(R)}(r)$  then take the form of \Eq{ELR}, in which
the coefficients are given by
\bea
C(k)&=&-\frac{\beta H(z)H'(n_r z)-H'(z)H(n_r z)}{\beta H(z)J'(n_r z)-H'(z)J(n_r z)}\,,
\label{Ccoef}
\\
B_1(k)&=&\frac{J(n_r z)H'(z)-\beta J'(n_r z)H(z)}{J(z)H'(z)-J'(z)H(z)}\,,
\nonumber\\
B_2(k)&=&-\frac{J(n_r z)J'(z)-\beta J'(n_r z)J(z)}{J(z)H'(z)-J'(z)H(z)}\,,
\nonumber\\
B_3(k)&=&\beta\frac{J(n_r z)H'(n_r z)-\beta J'(n_r z)H(n_r z)}{J(n_r z)H'(z)-\beta J'(n_r z)H(z)}\,,
\nonumber
\eea
where $z=kR$, $n_r$ and $\beta$ are defined in \Eq{nbe}, and the primes mean the derivatives of the functions with respect to their arguments.

Calculating the Wronskian \Eq{Wronskian}, we obtain
\be
W=\frac{n_r}{\mu} \left[ J(x)H'(x)-J'(x)H(x) \right]= i\beta\,,
\label{W}
\ee
using \Eq{nbe} and the Wronskian of the spherical Bessel equation~\cite{Abramowitz1964}:
\be
J(x)H'(x)-J'(x)H(x)= i\,.
\label{Wron}
\ee

\subsubsection{RS normalization}
\label{App:C2}

Let us first obtain \Eq{Az} for the normalization constant $A_n$, using the definition \Eq{Andef}. For this purpose, we Taylor expand the denominator $D(k)$ in the constant $C(k)$ given by \Eq{Ccoef}, up to first order about the point $k=k_n$:
\bea
D(k)&=&\beta H(z)J'(n_r z)-H'(z)J(n_r z)
\nonumber\\
&\approx& \beta [H(z_0)+H'(z_0)(z-z_0)]
\nonumber\\
&&\times[J'(n_r z_0)+J''(n_r z_0)n_r(z-z_0)]
\nonumber\\
&&- [H'(z_0)+H''(z_0)(z-z_0)]
\nonumber\\
&&\times[J(n_r z_0)+J'(n_r z_0)n_r(z-z_0)]
\nonumber\\
&=& (k-k_n)R \frac{H(z_0)}{J(n_r z_0)} \left\{J'^2(n_r z_0)\eps\left(\frac{1}{\mu}-1\right)\right.
\nonumber\\
&&
\left.+J^2(n_r z_0)\left[\frac{\alpha^2}{z_0^2}\left(\frac{1}{\mu}-1\right)+1-\eps\right]\right\}\,,
\label{Dkn}
\eea
where $z=kR$ and  $z_0=k_nR$. In doing so we have
used the secular equation (\ref{secular}) and Bessel's equation
\be
F''(z)=(\alpha^2/z^2-1)F(z)\,,
\label{Bess}
\ee
valid for $F(z)=J(z)$ or $F(z)=H(z)$ [compare with \Eqs{WE1}{WE2}].  The numerator in $C(k)$ is given by
\bea
N(k_n)&=&-\beta H(z_0)H'(n_r z_0)+H'(z_0)H(n_r z_0)
\nonumber\\
&=&-\beta \frac{H(z_0)}{J(n_r z_0)}
\nonumber\\
&& \times\left[J(n_r z_0)H'(n_r z_0)-J'(n_r z_0)H'(n_r z_0)\right]
\nonumber\\
&=&-i\beta \frac{H(z_0)}{J(n_r z_0)}\,,
\label{Nkn}
\eea
again using the Wronskian \Eq{Wron} and the secular equation (\ref{secular}). Substituting $D(k)$ and $N(k_n)$ from \Eqs{Dkn}{Nkn} into the definition \Eq{Andef} and taking the limit, we obtain the normalization constant \Eq{Az}.

The same result can be obtained from the general normalization \Eq{Norm}, or its spherically symmetric version \Eq{norm-rad}. The latter can be written as
\be
1=I_V+I_S\,,
\label{IVS}
\ee
where the volume integral $I_V$, for a homogeneous sphere, transforms into
\bea
I_V&=&\int_0^R (\eps \cE_1^2+\mu \cH_2^2+\mu\cH_3^2)dr
\nonumber\\
&=&\frac{\eps A_n^2}{n_r k_n} \int_0^{n_r z_0} \left[J^2(x)(1+\alpha^2/x^2)+J'^2(x)\right]dx
\nonumber\\
&=&\eps A_n^2 R \left[J^2(x)(1-\alpha^2/x^2)+J'^2(x)\right]_{x=n_rz_0}
\label{IV}
\eea
with $z_0=k_nR$, after integrating by parts and using \Eq{Bess} and the analytic integral \Eq{Ipp} given below.

For the surface term $I_S$, which is evaluated at point $r=R_+$ outside the sphere, we need to consider the RS wave function outside the system, which is given by
$$
\Ec_n(r)\equiv
\left(\begin{array}{c}
\cE_n(r)\\
\cK_n(r)\\
\cM_n(r)
\end{array}\right)
= B_n
\left(\begin{array}{c}
H(y)\\
- H'(y)\\
-\alpha H(y)/y
\end{array}\right)\,,
$$
similar to \Eq{EFn}, with $y=k_nr$ ($r>R$) and
\be
B_n=A_nJ(n_rz_0)/H(z_0)\,.
\label{Bn}
\ee
We then obtain
\bea
\!\!\!\!\!\!
I_S&=&  \frac{R}{k_n}\left.\left(\cK_n\frac{d\cE_n}{dr}-\cE_n\frac{d\cK_n}{dr}\right)\right|_{r=R_+}
\nonumber\\
&=&B^2_nR[-H'^2(z_0)+H(z_0)H''(z_0)]
\nonumber\\
&=&B^2_nR[-H'^2(z_0)+H^2(z_0)(\alpha^2/z^2_0-1)]
\nonumber\\
&=&A^2_nR\left[-\beta^2 J'^2(n_rz_0)+J^2(n_rz_0)(\alpha^2/z_0^2-1)\right]\!,\ \ \ \
\label{IS}
\eea
using \Eqs{secular}{Bn}. Substituting \Eqs{IV}{IS} into \Eq{IVS} we obtain the same normalization constant \Eq{Az}.

\subsubsection{Static pole of the dyadic GF}
\label{App:C3}

The static pole residue of the dyadic GF is given by two alternative forms \Eqs{R-sol1}{R-sol2}, in terms of
the scalar GFs $g$ and $\tilde{g}$, respectively. Let us find these GFs for a homogeneous sphere in vacuum. The first one has the following form
$$
g(r,r')=\frac{f_L(r_<)f_R(r_>)}{\cW}\,,
$$
where $f_L(r)$ and $f_R(r)$ are solutions of the differential equation
$$
\left(\frac{1}{r^2}\frac{d}{dr}r^2 \frac{d}{dr} - \frac{\alpha^2}{r^2}\right)f_{L,R}(r)=0
$$
satisfying, respectively, the left and right boundary conditions, $f_L(0)=f_R(\infty)=0$. Both solutions, $f_L(r)$ and $f_R(r)$, satisfy also the continuity conditions on the sphere surface  of $f(r)$ and $\mu(r)f'(r)$, where $f'=df/dr$. Therefore they take the following explicit form:
$$
\begin{array}{l}
f_L(r)=\left\{\begin{array}{ll}
(r/R)^l & \ r\leqslant R\\
\tilde{A} (r/R)^l+\tilde{B}(r/R)^{-l-1} & \ r>R\,,
\end{array}
\right.
\smallskip
\\
f_R(r)=
\left\{\begin{array}{ll}
\tilde{C} (r/R)^l+\tilde{D}(r/R)^{-l-1}  & r\leqslant R\\
(r/R)^{-l-1}  & r>R\,,
\end{array}
\right.
\end{array}
$$
where
\be
\tilde{A}=\mu \tilde{D}=\frac{\mu l+l+1}{2l+1}\,,\ \ \ \
\tilde{B}=-\frac{l}{l+1}\mu \tilde{C}=-\frac{(\mu-1)l}{2l+1}\,.
\label{const}
\ee
The Wronskian is given by
$$
\cW=\mu(r)r^2[f_L(r)f'_R(r)-f'_L(r)f_R(r)]=-(\mu l+l+1)R\,.
$$

The other scalar GF has a similar form:
$$
\tilde{g}(r,r')=\frac{\tilde{f}_L(r_<)\tilde{f}_R(r_>)}{\tilde{\cW}}\,,
$$
where $\tilde{f}_L(r)$ and $\tilde{f}_R(r)$ are solutions of the differential equation
$$
\left(\frac{d^2}{dr^2} - \frac{\alpha^2}{r^2}\right)\tilde{f}_{L,R}(r)=0
$$
with $\tilde{f}_{L,R}(r)$ and $\frac{1}{\mu(r)}\tilde{f}'_{L,R}(r)$ being continuous, in accordance with \Eq{g-equ2}. They also satisfy the individual conditions $\tilde{f}_L(0)=\tilde{f}_R(\infty)=0$. Therefore, they  take the following  form:
$$
\begin{array}{l}
\tilde{f}_L(r)=\left\{\begin{array}{ll}
(r/R)^{l+1} & \ r\leqslant R\\
\tilde{D} (r/R)^{l+1}+\tilde{C}(r/R)^{-l} & \ r>R\,,
\end{array}
\right.
\smallskip
\\
\tilde{f}_R(r)=
\left\{\begin{array}{ll}
\tilde{B} (r/R)^{l+1}+\tilde{A}(r/R)^{-l}  & r\leqslant R\\
(r/R)^{-l}  & r>R\,,
\end{array}
\right.
\end{array}
$$
where the constants $\tilde{A}$, $\tilde{B}$, $\tilde{C}$, and $\tilde{D}$ are given by \Eq{const}.
The Wronskian takes the form
$$
\tilde{\cW}=\frac{\tilde{f}_L(r)\tilde{f}'_R(r)-\tilde{f}'_L(r)\tilde{f}_R(r)]}{\mu(r)}=-\frac{\mu l+l+1}{\mu R}\,.
$$

The scalar GFs $g$ and $\tilde{g}$, the static pole residue, and the 4th ML representation following from it are then given by explicit expressions provided in \Sec{Sec:pole}.

\subsubsection{Static-mode sets}
\label{App:C4}

{\it Set 1.} This set of static modes is generated by the scalar GF $g(r,r')$. For LM modes, the static-mode potentials are given by
\be
\psi_\lambda(r)=\frac{\phi_\lambda(r)}{\lambda}\,,
\label{psiphi}
\ee
where $\phi_\lambda(r)$ are solutions of the differential equation (\ref{psi}) with $w(r)=\mu(r) r^2$, which for a homogeneous sphere simplifies to
\be
\left(\frac{1}{r^2}\frac{d}{dr} r^2\frac{d}{dr} - \frac{\alpha^2}{r^2}+ \lambda^2 \Theta(R-r)\right) \phi_\lambda(r)=0\,.
\label{set1equ}
\ee
The above equation has to be solved with the
boundary conditions of continuity of $\phi_\lambda(r)$ and $\mu(r)\phi'_\lambda(r)$. This results in wave functions
\be
\phi_\lambda(r)={\lambda}{A_\lambda}\left\{\begin{array}{ll}
j_l(\lambda r) & \ r\leqslant R\\
j_l(\lambda R) (r/R)^{-l-1} & \ r>R
\end{array}
\right.
\label{psi1}
\ee
and in a secular equation determining the eigenvalues $\lambda$:
\be
\lambda\mu R j_l'(\lambda R)+ (l+1)j_l(\lambda R)=0\,.
\label{sec1}
\ee
Using \Eq{psi1}, the normalization condition, given by \Eq{psi-norm}, reduces to
\be
1=\mu\int_0^R\phi^2_\lambda(r) r^2 dr=A_\lambda^2\frac{\mu}{\lambda}\int_0^{\lambda R} J^2(x)dx\,,
\label{Norm-set1}
\ee
which determines the normalization constants $A_\lambda$. The last integral has the analytic form
\Eq{Ipp} given below. The wave functions $\psi_\lambda(r)$ defined in this way through \Eq{psiphi} present the VC basis set introduced in~\cite{LobanovPRA19}. In terms of this basis, the scalar GF $g(r,r')$, contributing to the static mode pole of the dyadic GF via \Eq{R-sol1}, is expressed as
\be
g(r,r')=-\sum_\lambda \psi_\lambda(r)\psi_\lambda(r')\,,
\label{set1g}
\ee
using the expansion \Eq{G0} and the link \Eq{psiphi}.
\bigskip

{\it Set 2.} Using the explicit expressions \Eq{gg} for the scalar GFs $g$ and $\tilde{g}$ in the region within the sphere \mbox{$(r\leqslant R)$,} we find
\be
g(r,r')=\frac{\tilde{g}(r,r')}{\mu^2 r r'} -\frac{1}{\mu R}\,\frac{\mu-1}{\mu l+l +1} \xi(r)\xi(r')\,,
\label{ggtilde}
\ee
where $\xi(r)$ is defined by \Eq{xi-eta} [see also \Eqs{R22R33}{c-def}]. The GF $\tilde{g}$, satisfying \Eq{g-equ2}, has a series representation given by the general \Eq{G0}, now having the form
$$
\tilde{g}(r,r')=-\sum_\lambda\frac{\tilde{\phi}_\lambda(r)\tilde{\phi}_\lambda(r')}{\lambda^2}\,,
$$
where $\lambda$ satisfies another secular equation provided below. The static-mode functions $\tilde{\phi}_\lambda(r)$ are solutions of the differential equation
\be
\left(\frac{d^2}{dr^2}  - \frac{\alpha^2}{r^2}+ \lambda^2 \Theta(R-r)\right) \tilde{\phi}_\lambda(r)=0\,,
\label{set2equ}
\ee
which respect the boundary conditions of continuity of $\tilde{\phi}_\lambda(r)$ and $\frac{1}{\mu(r)}\tilde{\phi}'_\lambda(r)$, following from \Eq{psi} used for $w(r)=1/\mu(r)$ in the operator $\hcL(r)$. They have the following explicit form
$$
\tilde{\phi}_\lambda(r)={\mu}{A}_\lambda\left\{\begin{array}{ll}
J(\lambda r) & \ r\leqslant R\\
J(\lambda R) (r/R)^{-l} & \ r>R\,,
\end{array}
\right.
$$
where $\lambda$ is given by a new secular equation
\be
\lambda R j_l'(\lambda R)+ (\mu l+1)j_l(\lambda R)=0\,.
\label{sec2}
\ee
The normalization constants are again defined by the general equation (\ref{psi-norm}),
\be
1=\frac{1}{\mu}\int_0^R\tilde{\phi}^2_\lambda(r) dr={A}_\lambda^2\frac{\mu} {\lambda}\int_0^{\lambda R} J^2(x)dx\,,
\label{Norm-set2}
\ee
which results in exactly the same analytic expressions for ${A}_\lambda$ as given by \Eq{Norm-set1}.

Within the sphere $(r\leqslant R)$, a series representation of $g(r,r')$ contributing to the static pole via \Eq{R-sol1}, in terms of the potentials ${\psi}_\lambda(r)$ is given by
\be
g(r,r')=-\sum_\lambda {\psi}_\lambda(r){\psi}_\lambda(r') - {\psi}_0(r){\psi}_0(r')
\label{set2g}
\ee
with
$$
{\psi}_\lambda(r)={{A}_\lambda}  j_l(\lambda r)
$$
having the same form as in {\it Set 1}. The last term in \Eq{set2g} is described in terms of
$$
{\psi}_0(r)= \sqrt{\frac{1}{\mu R}\,\frac{\mu-1}{\mu l+l+1}}  \left(\frac{r}{R}\right)^l\,,
$$
which can be interpreted as an additional static mode with $\lambda=0$.
\bigskip

{\it Set 3.} We consider here one more set of static modes, called volume-surface charge (VSC) basis which was also introduced in~\cite{LobanovPRA19}. This set of modes corresponds to a rather extreme boundary condition which is
that the wave function is vanishing everywhere outside the system. The differential equation for these static modes is similar to \Eqs{set1equ}{set2equ}:
\be
\left(\frac{1}{r^2}\frac{d}{dr} r^2\frac{d}{dr} - \frac{\alpha^2}{r^2}+ \lambda^2\right) \bar{\phi}_\lambda(r)=0\,,
\label{set3equ}
\ee
this time lacking any Heaviside function. In fact,
it needs to be solved only within a finite interval $0\leqslant r\leqslant R$ with the boundary condition $\bar{\phi}_\lambda(R)=0$, leading to the most simple secular equation
\be
j_l(\lambda R)=0\,.
\label{sec3}
\ee
The wave functions of the static modes are given  by
\be
\psi_\lam(r)= \frac{\bar{\phi}_\lambda(r)}{\sqrt{\mu}\lam}={A}_\lambda j_l(\lambda r)\,,
\label{psiphi3}
\ee
and the normalization constants ${A}_\lambda$, again determined by \Eq{psi-norm}, this time with $w(r)=r^2$, are thus taking  the form
\be
1=\int_0^R\bar{\phi}^2_\lambda(r) r^2 dr={A}_\lambda^2 \frac{\mu}{\lambda}\int_0^{\lambda R} J^2(x)dx\,,
\ee
identical to \Eqs{Norm-set1}{Norm-set2}.

In order to use this set of static modes, let us introduce a scalar GF $\bar{g}(r,r')$ corresponding to the problem described by \Eq{set3equ}. It satisfies a differential equation
\be
\left(\frac{1}{r^2}\frac{d}{dr} r^2\frac{d}{dr} - \frac{\alpha^2}{r^2}\right) \bar{g}(r,r')=\frac{\delta(r-r')}{ r^2}
\label{equ-set3}
\ee
and vanishing boundary conditions,
$\bar{g}(R,r')=\bar{g}(r,R)=0$. Solving \Eq{equ-set3} with the help of the left and right functions,
\bea
\bar{g}(r,r')&=&\frac{\bar{f}_L(r_<)\bar{f}_R(r_>)}{\bar{W}}\,,
\nonumber\\
\bar{f}_L(r)&=&(r/R)^l\,,
\nonumber\\
\bar{f}_R(r)&=&-(r/R)^l+(r/R)^{-l-1}\,,
\nonumber\\
\bar{W}&=&r^2\left[\bar{f}_L(r)\bar{f}'_R(r)-\bar{f}_L'(r)\bar{f}_R(r)\right]=-(2l+1)R\,,
\nonumber
\eea
we find
\be
\bar{g}(r,r')=\frac{1}{(2l+1)R}\xi(r)\xi(r') - \frac{1}{(2l+1)R}\xi(r_<)\eta(r_>)\,,
\label{gbar}
\ee
where functions $\xi(r)$ and $\eta(r)$ are defined by \Eq{xi-eta}.
On the other hand, $\bar{g}(r,r')$ has the static-mode representation,
\be
\bar{g}(r,r')=-\sum_\lambda\frac{\bar{\phi}_\lambda(r)\bar{\phi}_\lambda(r')}{\lambda^2}\,,
\label{gbar2}
\ee
according to \Eq{G0}. Comparing \Eqs{gbar}{gg}, and using the series \Eq{gbar2} and the relation \Eq{psiphi3}, we find
\be
g(r,r')=-\sum_\lambda {\psi}_\lambda(r){\psi}_\lambda(r') - {\psi}_0(r){\psi}_0(r')\,,
\label{set3g}
\ee
where
$$
{\psi}_0(r)= \sqrt{\frac{1}{R}\,\frac{1}{\mu l+l+1} } \left(\frac{r}{R}\right)^l\,.
$$
\bigskip

{\it Discussion.} The static-mode sets considered above clearly demonstrate a flexibility of their choice for ML representations and the RSE. Comparing all three sets of static modes presented, we see that they provide alternative series representations of the scalar GF $g(r,r')$. All three representation have the same form except that the series \Eq{set1g} for Set 1 is lacking the $\lambda=0$ term, as compared to \Eqs{set2g}{set3g} which are identical. This term can be formally introduces for Set 1 as well, by defining a vanishing amplitude of the function $\psi_0(r)$ for this set. The normalization constants in $\psi_0(r)$ then take different analytic form among all three sets.

In all three sets of static modes, the basis wave functions (with $\lambda\neq 0$) have exactly the same form in terms of  $\lambda$,
\be
\psi_\lam(r)= {A}_\lambda j_l(\lambda r)\ \ (r\leqslant R)
\ee
with the normalization constants $A_\lam$ given by the following explicit expressions~\cite{LobanovPRA19}:
\be
A_\lam^2=\frac{2}{\mu\lambda^2 R^3}\left[  j^2_l(\lambda R)-j_{l-1}(\lambda R)j_{l+1}(\lambda R)\right]^{-1}\,.
\ee
The eigenvalues $\lambda$ are different for different static-mode sets and are determined by the secular equations (\ref{sec1}), (\ref{sec2}), and (\ref{sec3}), following from different boundary conditions on the sphere surface, imposed for the static-mode wave functions.  Consequently, the actual values of ${A}_\lambda$ and the actual wave functions $\psi_\lam(r)$ are also different.

Using the relation \Eq{R-sol1} between the static pole of the dyadic GF and the scalar GF $g(r,r')$, one can find from \Eqsss{set1g}{set2g}{set3g} the static-pole part of the 1st ML representation \Eq{MLrep1}. In that representation, the vector functions $\Ec_\lambda(r)$ are defined by \Eq{static1} with the scalar fields $\psi_\lambda(r)$ generated above for each set of modes. Note that when using \Eq{MLrep1} for Sets 2 and 3, the series should include a $\lambda=0$ term due to the additional  effective static mode contributing to these sets, as discussed above. The 3rd ML representation \Eq{ML4} can be obtained from the above series for $g$, by using the relation between $g$ and $\cG_{22}$ provided by \Eq{R22g}. Finally, the effective $\lam=0$ mode of Set 2 contributes to the last term of the 4th ML representation \Eq{ML5}, in which all the physical static modes have been eliminated.


\subsubsection{Matrix elements}
\label{App:C5}

Let us introduce for convenience the following analytic vector function
$$
\Ec(k,r)\equiv
\left(\begin{array}{c}
\cE(k,r)\\
\cK(k,r)\\
\cM(k,r)
\end{array}\right)
= \cA(z)
\left(\begin{array}{c}
J(x)\\
-\beta J'(x)\\
-\alpha\beta J(x)/x
\end{array}\right)\,,
$$
where $x=n_r k r$, $z=n_r k R$, and the normalization function $\cA(z)$ is defined by \Eq{Az}. Clearly,
$\Ec_n(r)=\Ec(k_n,r)$ at the RS wave numbers $k_n$, see \Eq{EFn}.

Looking at the ML representation \Eq{ML5}, it is easy to see that only the integrals of the four products
$\cE(p,r)\cE(q,r)$, $\cK(p,r)\cK(q,r)$, $\cM(p,r)\cM(q,r)$, and $\cK(p,r)\cM(q,r)$ contribute to all possible matrix elements for a spherical shell perturbation, three of them being already outlined in \Eq{Vnm}. This implies that the matrix elements can be expressed in terms of the following integrals of spherical Bessel functions:
\bea
I_1(p,q)&=&\int_{R_1}^{R_2} J(pr)J(qr)dr\,,
\nonumber\\
J_1(p,q)&=&\int_{R_1}^{R_2} J'(pr)J'(qr)dr\,,
\nonumber\\
J_2(p,q)&=&\int_{R_1}^{R_2} \frac{J(pr)}{pr}\frac{J(qr)}{qr}dr\,,
\nonumber\\
J_3(p,q)&=&\int_{R_1}^{R_2} J'(pr)\frac{J(qr)}{qr}dr\,.
\nonumber
\eea
$I_1(p,q)$ is a well-known analytic integral, which is given by
$$
I_1(p,q)=\left. \frac{qJ(pr)J'(qr)-pJ'(pr)J(qr)}{p^2-q^2}\right|_{R_1}^{R_2}
$$
for $p\neq q$, and by
\be
I_1(p,p)\!=\!\frac{1}{2p}\!\left[z\!\left\{\!J^2(z)\!\left(\!1\!-\!\frac{\alpha^2}{z^2}\right)\!
+\!J'^2(z)\!\right\}\!-\!J(z)J'(z)\right]_{\!pR_1}^{\!pR_2}
\label{Ipp}
\ee
for $p=q$.
Integrals $J_1(p,q)$, $J_2(p,q)$, and $J_3(p,q)$, when considered separately, have to be evaluated numerically. Their combinations, however, produce another analytic integral:
$$
I_2(p,q)=J_1(p,q)+\alpha^2J_2(p,q)\,,
$$
which is given by
$$
I_2(p,q)=\left. \frac{pJ(pr)J'(qr)-qJ'(pr)J(qr)}{p^2-q^2}\right|_{R_1}^{R_2}
$$
for $p\neq q$, and by
$$
I_2(p,p)\!=\!\frac{1}{2p}\!\left[z\!\left\{\!J^2(z)\!\left(\!1\!-\!\frac{\alpha^2}{z^2}\right)\!
+\!J'^2(z)\!\right\}\!+\!J(z)J'(z)\right]_{\!pR_1}^{\!pR_2}
$$
for $p=q$. Also, the following analytic integral may serve for verification of the numerics:
$$
I_3(p,q)= J_3(p,q)+J_3(q,p)-J_2(p,q)
=\left. \frac{J(pr)J(qr)}{pqr}\right|_{R_1}^{R_2}\,.
$$


\end{document}